\documentclass[11pt,a4]{article}

\usepackage{rotating}
\usepackage{citesort}
\newcommand{\smf}[2]{\hspace{-1.8mm}\begin{array}{c} \\[-5.5mm] \frac{#1}{#2}\end{array}\hspace{-1.8mm}}
\newcommand{\hf}{\smf{1}{2}}
\newcommand{\roemII}{I\hspace{-0.5mm}I}
\newcommand{\mbf}{\mathbf}

\begin{document}
\title{\bf {Nonzero angular momentum states of the helium atom in a strong magnetic field}}

\author{W.Becken and P.Schmelcher\\
Theoretische Chemie\\
Physikalisch-Chemisches Institut\\
Im Neuenheimer Feld 229\\69120 Heidelberg\\Federal Republic of Germany}

\date{}
\maketitle

\begin{abstract}
The electronic structure of the helium atom in the magnetic field regime $B=0-100a.u.$ is 
investigated, using a full configuration interaction approach which is based on a nonlinearly 
optimized anisotropic Gaussian basis set of one-particle functions. The corresponding generalized 
eigenvalue problem is solved for the magnetic quantum number $M=-1$ and for both even and odd 
$z$-parity as well as singlet and triplet spin symmetry. Accurate total electronic energies of the 
ground state and the first four excitations in each subspace as well as their one-electron 
ionization energies are presented as a function of the magnetic field. Additionally we present 
energies for electromagnetic transitions within the $M=-1$ subspace and between the $M=-1$ subspace 
and the $M=0$ subspace treated in a previous work. A complete table of wavelengths and field 
strengths for the detected stationary points is given. 
\end{abstract}
\vspace{1cm}

\section{Introduction}

Motivated by the astrophysical discovery of strong magnetic fields on the surfaces of white dwarfs 
($\le 10^{5}$ Tesla) and neutron stars ($\approx 10^{8}$ Tesla), the  behaviour and the properties 
of matter in strong magnetic fields has increasingly attracted interest. The theoretical description 
of atoms in strong magnetic fields is well covered in the literature only for the case of the 
hydrogen atom (see refs.\cite{friedrich,roesner,wintgen,ivanov88,kravchenko,ruder,schmelchschweiz} 
and refs. therein). Until recently our knowledge about atoms with more than one electron in a strong 
magnetic field has been relatively sparse and definitely not sufficient for a comparison of the 
corresponding theoretical data with the mysterious absorption edges \cite{green,schmidt,schmidt96} 
in the spectrum of the magnetic white dwarf GD229. Those have for the first time been identified as 
helium lines by Jordan {\it et al}\cite{jssbecken} in 1998. This overwhelming evidence was based on 
results of highly accurate {\it ab initio} calculations performed by Becken and Schmelcher in 1998, 
a part of which has been published in ref.\cite{helium}. For a more detailed overview over the 
various theoretical approaches to the helium atom in a strong magnetic field and the corresponding 
literature before 1998 we refer the reader to \cite{helium} and in particular the references 
therein. In \cite{helium} a fully correlated configuration interaction approach has been applied to 
the helium atom in a strong magnetic field. Total energies for spin singlet and triplet states for 
both positive and negative $z$-parity in the subspace of vanishing magnetic quantum number $M=0$ 
have been provided, thereby covering the regime of magnetic fields strengths from $B=0$ to 
$B=100a.u.$ ($B=1a.u.$ corresponds to $2.35\cdot 10^5 $Tesla). Additionally all the transition 
energies within the $M=0$ subspace have been presented and discussed there, including the stationary 
components with respect to the field dependence which have been the key tool for the comparison with 
the observed spectra \cite{jssbecken}.

The aim of the present paper is to provide important data for states with the magnetic quantum 
number $M=-1$. Some data for states with positive $z$-parity and $M=-1$ have already successfully 
been used for a comparison with the astronomical observation(in \cite{jssbecken}, we used the 
corresponding stationary transitions in a certain field and wavelength regime). The {\it complete} 
data are presented here for the first time. Data resulting from states with negative $z$-parity are 
very recent results. The latter further extend our knowledge on the helium atom in a strong magnetic 
field and permit to investigate additional transitions.

For the case of triplet spin symmetry, Jones {\it et al} \cite{jones99} very recently used a 
released-phase quantum Monte Carlo method for calculating accurate data. However, they cover only 
three field strengths, investigate less excited states for common symmetries and do not study the 
spin singlet states at all. Nevertheless for the common values of the field strength they confirm 
our data to several digits. In contrast to this the energies presented by Scrinzi\cite{scrinzi}, 
though in principle variational, are in the presence of a magnetic field significantly lower than 
ours. Carefully comparing the energies, it appears to us that they are systematically too low, i.e. 
most probably due to a numerical error.

We remark that performing calculations for finite magnetic quantum numbers requires within our 
approach drastically improved computational techniques (in comparison to the case $M=0$) for keeping 
the CPU time affordable. A summary of these techniques is provided in the appendix of the present 
paper. 

The starting point of the present paper is the nonrelativistic Hamiltonian of the helium atom with 
infinite nuclear mass in a magnetic field as given in section 2. To be self-contained we briefly 
discuss the Hamiltonian's symmetries and provide a description of the basis set as well as the full 
configuration interaction approach (for more details see ref.\cite{helium}). We introduce a maximal 
set of conserved quantities, chosen to be the total spin $S^2$, the $z$-component $S_z$ of the total 
spin, the total spatial magnetic quantum number $M$ and the total spatial $z$-parity $\Pi_z$. These 
symmetries serve for classifying the results for the energies for $M=-1$ in sec.3. In each of the 
two subspaces for positive and negative $z$-parity we present the total energies and the ionization 
energies of the ground state and the first four excited states for singlet and triplet spin 
symmetry. Additionally we consider in sec.3 all the transitions within the $M=-1$ subspace as well 
as all the possible transitions between the $M=-1$ states treated in the present paper and the $M=0$ 
states given in ref.\cite{helium}. The wavelengths of all the stationary components are provided, 
being the basic ingredient for the successful comparison of theoretical data with the spectrum of 
magnetic white dwarfs in general and specifically for GD229.

\section{Hamiltonian, Symmetries and basis sets}\label{sec_ham_bas}

\subsection{Hamiltonian and Symmetries}

Assuming the magnetic field to point in the $+z$-direction, the Hamiltonian reads
\begin{eqnarray}
	H  &=&   \sum\limits_{i=1}^2 \left(   \hf\mbf{p}_i^2 +\hf B{{l}_z}_i 
				            + \smf{B^2}{8}(x_i^2+y_i^2) 
		                            - \smf{2}{|\mbf{r}_i|} 
                                            + B{{s}_z}_i 
				     \right)
		 \;\;+\;\;   \frac{1}{|\mbf{r}_2 - \mbf{r}_1|}                           \label{ham}
\end{eqnarray}
The one-particle operators in eq.(\ref{ham}) are the Coulomb potential energies 
$-\smf{2}{|\mbf{r}_i|}$ of the electrons in the field of the nucleus as well as their kinetic 
energies, here splitted into the parts $\hf \mbf{p}_i^2$, the Zeeman terms $\hf B{{l}_z}_i $, the 
diamagnetic terms $\smf{B^2}{8}(x_i^2+y_i^2)$ and their spin energies $B{{s}_z}_i$. The 
electron-electron repulsion energy is represented by the two-particle operator 
$\frac{1}{|\mbf{r}_2 - \mbf{r}_1|}$. We remark that we use an electron spin $g$-factor equal 2, and 
any more accurate value for it can be simply incorporated by shifting the final total energies 
correspondingly. For remarks on the influence of relativistic effects and on a scaling relation 
taking into account the finite nuclear mass we refer the reader to ref.\cite{ruder,helium,pavlov}.

Analogously to ref.\cite{helium}, we exploit that there exist four independent commuting conserved 
quantities: the total spin $\mbf{S}^2$, the $z$-component $S_z$ of the total spin, the 
$z$- component $L_z$ of the total angular momentum and the total spatial $z$-parity $\Pi_z$. 

\subsection{Basis sets}

For constructing a two-particle basis set of eigenfunctions of the above mentioned conserved 
quantities, our central ingredient is an anisotropic Gaussian basis set of one-particle functions
\begin{eqnarray}
	\Phi_i(\rho,\varphi,z)  &=&     \rho^{{n_\rho}_i} z^{{n_z}_i} 
                                     e^{-\alpha_i\rho^2-\beta_i z^2} 
                                     e^{im_i\varphi}  \;\;\;\;i=1,...,n\;\;\;,        \label{onebas}
\end{eqnarray}
which are themselves eigenfunctions of the corresponding one-particle operators of the mentioned 
conserved quantities. The parameters ${n_\rho}_i$ and ${n_z}_i$ are restricted by
\begin{eqnarray}
     {n_\rho}_i  &=& |m_i|     +  2k_i\;;\;\;\; k_i = 0,1,2,...\;\;with\;\;m_i = ...-2,-1,0,1,2,...
										      \label{gerade}
                                                                                                  \\ 
     {n_z}_i     &=& \;{\pi_z}_i\; +  2l_i\;;\;\;\;\; l_i = 0,1,2,...\;\;with\;\;{\pi_z}_i = 0,1 
                                                                                        \label{zpar}
\end{eqnarray}
whereas the nonlinear variational parameters $\alpha_i$ and $\beta_i$ are positive and have to be 
nonlinearly optimized for each field strength as described in ref.\cite{helium}. For each 
one-particle subspace of given symmetry we used an algorithm for determining the nonlinear 
parameters $\alpha_i$ and $\beta_i$ such that the states of the hydrogen atom or the He$^+$ ion for 
that symmetry were optimally described. We emphasize that this procedure gives rise to considerable 
effort since it has to be repeated for each field strength separately.

We construct a basis set of spatial two-particle states by
\begin{eqnarray}
	\left.|\psi_q\right\rangle &:=& b_i^{\dag} b_j^{\dag}\left. |0\right\rangle   
		\;\;\;\;i=1,...,n\;\;\;,\;\;\;\;j=i,...,n\;\;\;,                      \label{twobas}
\end{eqnarray}
where $b_i^{\dag}$ is the creation operator of the $i$-th one-particle state 
$\left.|i\right\rangle = b_i^{\dag}\left.|0\right\rangle$ whose position representation is given by 
eq.(\ref{onebas}). The spin space is spanned by spin singlet or spin triplet states, and therefore 
the operators $b_i^{\dag}$ have to be chosen bosonic or fermionic, respectively. Selecting 
combinations with 
\begin{eqnarray}
	             m_i+m_j &=& M        \;\;\;\;,\;\;\;\;\;\;\;\;\;
                     mod({\pi_z}_i + {\pi_z}_j,2) \;\;=\;\; \Pi_z  \;\;\;,
\end{eqnarray}
we achieve the two-particle states (\ref{twobas}) to be a basis set within the subspace for given 
total symmetries $M$ and $\Pi_z$. The number $N$ of two-particle basis states is thus in general 
smaller than $n(n+1)/2$.

We perform a {\bf full Configuration Interaction (full CI)} approach by representing the Hamiltonian 
in a basis whose spatial part is given by the in general nonorthonormal states (\ref{twobas}). Since 
the spin part $B\sum {{s}_z}_i$ of the Hamiltonian can trivially be taken into account by a shift of 
the energies it is sufficient to represent the spatial part of the Hamiltonian $H$ and the overlap 
$S$ by 
\begin{eqnarray}
         S_{pq}    &=&       \left\langle \psi_p|  \psi_q \right\rangle\;\;,\;\;\;\;\;   
         H_{pq} \;\;=\;\;    \left\langle \psi_p|H|\psi_q \right\rangle 
\end{eqnarray}
The matrices $S$ and $H$ are Hermitian, and the overlap $S$ is additionally positive definite. 
Furthermore the matrix elements turn out to be real. The finite-dimensional generalized 
real-symmetric eigenvalue problem
\begin{eqnarray}
	(\underline{\underline{H}} - E \underline{\underline{S}})\cdot \underline{c}  &=& 0     
										   \label{seculareq}
\end{eqnarray}
provides eigenvalues $E$ which are variational upper bounds to the exact eigenvalues of the 
Hamiltionian (\ref{ham}) within each subspace of given $M$ and $\Pi_z$.

\subsection{Matrix elements}

For calculating the matrix elements of the spatial part of the Hamiltonian (\ref{ham}), we rewrite 
the former in second quantization, $\hat{H}  =  \hat{H}_{I}  +  \hat{H}_{\roemII}$, where 
$\hat{H}_{I}$ and $ \hat{H}_{\roemII}$ denote the second-quantized counterparts of the familiar one- 
and two-particle operators whose position representations read
\begin{eqnarray}
	H_{I}(\mbf{p},\mbf{r})            &=&      \hf \mbf{p}^2 + \hf \mbf{B}\cdot\mbf{l} 
                                                 + \smf{1}{8}B^2(x^2+y^2)  - \smf{2}{|\mbf{r}|}
                                                                                      \hspace{1.6cm}
	H_{\roemII}(\mbf{r}_1,\mbf{r}_2)  \;\;=\;\;      \smf{1}{|\mbf{r}_2 - \mbf{r}_1|} 
\end{eqnarray}
Now, with 
    $\left.|\psi_q\right\rangle := b_i^{\dag} b_j^{\dag}\left. |0\right\rangle$ 
and $\left.|\psi_p\right\rangle := b_k^{\dag} b_l^{\dag}\left. |0\right\rangle$ a straightforward 
calculation leads to
\begin{eqnarray}
        \left\langle \psi_p | \psi_q \right\rangle \;\;\;            
&=&     \;\;\;\;\;\;  \left\langle i | k \right\rangle \;\;\left\langle j | l \right\rangle \;\;\;\;\,
     \pm    \;\;\;\;  \left\langle i | l \right\rangle\;\; \left\langle j | k \right\rangle    
											\label{matrel1}\\
	\left\langle \psi_p |\hat{H}_{I}| \psi_q \right\rangle       
&=&           \;\;\;  \left\langle i | H_{I} | k \right\rangle   \; \left\langle j | l \right\rangle  
     \;\;\; \pm \;\,  \left\langle i | H_{I} | l \right\rangle \;   \left\langle j | k \right\rangle  
											     \nonumber \\
&&	    +         \left\langle j | H_{I} | l \right\rangle \;   \left\langle i | k \right\rangle 
     \,\;\; \pm  \;   \left\langle j | H_{I} | k \right\rangle \;   \left\langle i | l \right\rangle   \\ 
        \left\langle \psi_p |\hat{H}_{\roemII}| \psi_q \right\rangle 
&=&     \;\;\;\;\;\;  \left\langle ij | H_{\roemII} | kl \right\rangle \;\;\;\;
	\pm \;\;\;\;  \left\langle ij | H_{\roemII} | lk \right\rangle  \;\;\;\;,         \label{matrel3}
\end{eqnarray}
where $\left.|ij\right\rangle := \left.|i\right\rangle \otimes \left.|j\right\rangle$ and where the 
sign '$\pm $' stands for '$+$' in the singlet case and for '$-$' in the triplet case.

For the relatively simple evaluation of the $n(n+1)/2$ different one-particle overlaps 
$\left\langle i | k \right\rangle$ and matrix elements $\left\langle i | H_{I} | k \right\rangle$ we 
refer the reader to appendices A,B in ref.\cite{helium}. The two-particle matrix elements 
$\left\langle ij | H_{\roemII} | kl \right\rangle$ are by no means trivial, in particular in view of 
the fact that their accurate and fast evaluation is necessary in order to build up the Hamiltonian 
matrix in an affordable amount of CPU time. In ref.\cite{helium} we discussed a method using a 
decomposition in Cartesian coordinates which expresses the two-particle matrix elements in series of 
hypergeometric functions whose evaluation has been performed by highly efficient analytical 
continuation formulas. The latter are necessary in order to keep the CPU time acceptable since the 
number of different two-particle matrix elements is of the order $N(N+1)/2$ rather than $n(n+1)/2$. 
However, the Cartesian decomposition becomes more and more inefficient with increasing magnetic 
quantum number, which is already relevant for calculations of the subspace $M=-1$. Therefore we have 
developped a drastically improved procedure using cylindrical coordinates which leads to an enormous 
gain of speed such that the computation of the whole Hamiltonian matrix becomes even faster than its 
diagonalization by standard library routines. The derivation of the corresponding powerful formula 
for the electron-electron integral is rather lengthy and complicated: we therefore present only 
major steps of it in appendices A,B,C of the present paper.

\section{Results and discussion}

Throughout the paper we use the notation $\nu^{2S+1}_{S_z}M^{(-1)^{\Pi_z}}$ for a state with spin 
multiplicity $(2S+1)$ and degree of excitation $\nu =1,2,3,...$ within the subspace of given 
magnetic quantum number $M$ and $z$-parity $\Pi_z$. The index $S_z$ will be omitted in obvious 
cases. The present paper is concerned with the subspaces ${}^1(-1)^+$, ${}^3(-1)^+$, 
${}^1(-1)^-$, ${}^3(-1)^-$. The correspondence between our field notation and the common 
spectroscopic notation $n^{2S+1}_{S_z}L_M$ in field-free space is discussed in ref.\cite{helium} 
(see table 1 therein). For completeness we mention that the $5^1(\pm1)^-$ states correspond to the 
$6^1D_{\pm 1}$ states whose field free energy is $-2.138982274 a.u.$, and analogously the 
$5^3(\pm1)^-$ states correspond to the  $6^3D_{\pm 1}$ states with a field free energy of 
$-2.013901415 a.u.$ (Both values are taken from ref.\cite{drake}).

\subsection{Aspects for the selection of basis functions}

For the $M=0$ states treated in ref.\cite{helium}, we have been able to achieve a considerable 
accuracy by choosing basis sets which can describe the shape of the exact wave function, i.e. 
include electronic correlation effects. The latter become less important with increasing quantum 
number $|M|$, and this manifests itself already for $M=-1$. The reason is that bound two-particle 
states with nonzero values of $M$ are approximately one-particle excitations. Consequently, the 
electrons are spatially more separated than in a $0^+$ state, and this lowers the correlation 
energy. Additionally, the cusp problem is also less important for excited states $M=-1$, and 
therefore fewer one-particle functions with large values for the nonlinear $\alpha$ and $\beta$ 
parameters are needed. We have exploited these facts and achieved even more accurate results for 
the $M=-1$ states than for the $M=0$ states.
  
In detail the strategy was similar to the $M=0$ case. In the case of the $(-1)^+$ subspace we used 
260 optimized one-particle basis functions (for each field strength) for constructing a two-particle 
basis set of dimension $N=3793$. The latter number is identical for the singlet and the triplet 
subspace since it cannot occur that a two-particle state contains two identical one-particle 
contributions which combine to the odd total magnetic quantum number $M=-1$. This is a principal 
difference to the case of the $0^+$ subspace. In order to describe angular correlation, we have 
added also $(-2)^+$ and $(-3)^+$ functions which are paired with $(+1)^+$ and $(+2)^+$ functions, 
respectively, to build up $M=-1$. The same scheme was used for the contributions of one-particle 
functions of negative $z$-parities. There it was sufficient to use the combinations $(-1)^-/0^-$ and 
$(-2)^-/(+1)^-$. In order to describe excitations we added one-particle basis functions with quantum 
numbers $m^{\pi_z} = (-1)^+$ and values $l_i=1$ and $k_i=1$ (see eqs.(\ref{gerade},\ref{zpar})). The 
latter have exclusively been optimized for a nuclear charge number $Z=1$, in contrast to all the 
other types of basis functions which have been optimized for $Z=1$ (hydrogen) and for $Z=2$(He$^+$). 
The reason is again that all bound $M=-1$ states are one-particle excitations in which the excited 
electron is associated to the one-particle quantum numbers $m^{\pi_z} = (-1)^+$, and the effect of 
the nucleus on that outer electron is screened by the inner one, thereby giving rise to an effective 
nuclear charge close to unity.

For the $(-1)^-$ subspace we proceeded in a similar manner, the basis set dimension was $N= 3671$ 
for both singlet and triplet states, built up from a set of 228 optimized one-particle basis 
functions of type (\ref{onebas}).

In the following, we present and discuss the results for the helium energy calculations. Comparing 
our field-free data with the literature we observe that they are for the subspace $(-1)^+/(-1)^-$ 
subspaces more accurate than in the case of the $0^+/0^-$ subspaces. We will as far as possible 
compare our helium energies for finite field strength with the best data available in the 
literature.
\vspace{2mm}

\subsection{Energies for finite field strengths}

\subsubsection{Results for $M=-1$ and even $z$-parity}\label{m1p}

{\bf a) The singlet states $\nu^1(-1)^+$}

For the singlet subspace $\nu^1(-1)^+$ we present the energies of the ground state and the first 
four excitations, i.e. $1\le \nu \le 5$. The energies for the $1^1(-1)^+$ state are presented in 
table 1, together with the values given in the literature, if available. The accuracy of the field 
free energy is even higher than the ones of the $1^10^+$ and $1^10^-$ states due to the lower 
correlation energy for the $M=-1$ states. We remark that even though the data for the $(-1)^+$ 
symmetry presented in ref.\cite{jssbecken} have been accurate enough for identifying features in the 
spectrum of the magnetic white dwarf GD229 as absorption edges of helium, we have been able to 
improve the accuracy slightly.

We observe that for small values of $B$ the total energies of the $1^1(-1)^+$ state lie slightly 
lower than for $B=0$ which is an effect due to the Zeeman energy which is negative for $M=-1$. 
However, for fields stronger than $B\approx 0.16a.u.$ the total energies rise drastically with 
increasing field strength which has its origin in the increasing kinetic energy of the electrons 
which can roughly be estimated by their Landau energy amounting to $B$ for both electrons together.

In order to reveal the internal energetics of the atom we have to subtract such pure and overall 
field effects. For an analysis it is advantageous to subtract even more: one measure for the 
accuracy of the total energies $E(B)$ are their corresponding one-particle ionization energies 
$|E(B)-T(B)|$ corresponding to the process He $\rightarrow$ He$^+ + e^-$. The threshold $T(B)$, i.e. 
the lowest possible total energy for which the system He$^+ + e^-$ can exist possessing the {\it 
same} quantum numbers as the He state in question, is given in the fourth column in table 1 (the 
values for the ionization energies are trivial to compute and are not additionally listed in any of 
the tables). The values $T(B)$ for $(-1)^+$ symmetry are identical to those given in 
ref.\cite{helium} for the $M=0$ symmetry. The reason is that for any $z$-parity and spin symmetry 
the lowest $M=-1$ state of the system He$^+ + e^-$ is realized by the ionized electron in a Landau 
state with magnetic quantum number $m=-1$ and the He$^+$ ion in its $0^+$ ground state: the Landau 
energy of the electron depends on $(m+|m|)$ and is therefore identical for $m=-1$ and $m=0$. The 
alternative possibility of associating the value $m=-1$ to the He$^+$ ion possesses a higher 
energy. 

The energies for the excited states $\nu^1(-1)^+$, $2\le \nu \le 5$ are given in table 2. We remark 
that for finite field strengths there are no data about these states available in the literature so 
far. We observe as expected that their total energies rapidly approach the threshold $T(B)$ with 
increasing excitation.

The dependence of the one-particle ionization energies on the field strength is shown in figure 1. 
In contrast to the total energies of the $1^1(-1)^+$ state its ionization energy depends 
monotonously on $B$: the outer electron becomes increasingly bound with increasing field strength. 
This is not in general the case for the excited states within the $(-1)^+$ subspace although the 
ionization energy of the $2^1(-1)^+$ state is monotonous (note the changes in the slope compared to 
the $1^1(-1)^+$ state). The ionization energies of the states $3^1(-1)^+$, $4^1(-1)^+$ and 
$5^1(-1)^+$ exhibit a pattern of several avoided crossings. Those occur in the field strength 
interval $0.02 \le B \le 0.2$ which is the regime where a rearrangement takes place: for low field 
strengths the states $3^1(-1)^+$ and $4^1(-1)^+$ (field free $4^1F_{-1}$ and $4^1P_{-1}$) are 
energetically almost degenerate since they both belong to the field free principal quantum number 
$n=4$. This degeneracy is disturbed by the magnetic field and completely destroyed for strong 
fields (see figure 1).
\vspace{3mm}

\noindent
{\bf b) The triplet states $\nu^3(-1)^+$}

For the triplet subspace $\nu^3(-1)^+$ we present the energies for the ground state and the first 
four excitations, i.e. $1\le \nu \le 5$. The total energies of the states with $S_z=-1$ are given in 
table 3 together with data existing in the literature for the states $1^3(-1)^+$, $2^3(-1)^+$ and
$3^3(-1)^+$. The states $4^3(-1)^+$ and $5^3(-1)^+$ have not been present in the literature so far. 
We observe that our energies are variationally lower than any values in the literature apart from 
only a few exceptions with respect to the energies computed by Jones {\it et al} \cite{jones99}. 
However, one must take into account that the released-phase quantum Monte Carlo method performed
by Jones {\it et al} leads to statistical error bars (see the corresponding numbers in parentheses 
in table 3) such that none of those energies is significantly stronger bound than our results. This 
means, either we confirm these results or our accuracy is even higher, reflecting the fact that our 
optimized anisotropic Gaussian basis sets excellently describes the wave functions in a magnetic 
field. Due to the efficiency of our method of computation we were able to cover a large number of 
field strengths.

Due to the spin shift $BS_z$ the triplet states with $S_z=-1$ are the most strongly bound ones among 
the three states with $S_z=0,\pm 1$. This spin shift causes the $1^3_{-1}(-1)^+$ state to cross the 
low-field ground state $1^10^+$ at $B\approx 0.750a.u.$ For $B\stackrel{>}{\sim}0.750a.u.$ the 
$1^3_{-1}(-1)^+$ state is the global ground state of the atom. The $1^3_{-1}(-1)^+$ state is for any 
field strength lower than the $1^3_{-1}(-1)^-$ state (see sec. \ref{m1m}), and it lies also lower 
than the $1^3_{-1}0^+$ and $1^3_{-1}0^-$ states for $B \ge 0.750a.u.$. The latter ones cross the 
$1^10^+$ state at $B\approx 1.112a.u.$ and $B\approx 0.994a.u.$, respectively.

Analogously to the singlet case, we show the one-particle ionization energies of the triplet states 
also in figure 1. We observe that the singlet-triplet splitting decreases with increasing 
excitation. This occurs due to the fact that for excitations the spatial separation between the 
electrons is large, and therefore the exchange terms in eqs.(\ref{matrel1}-\ref{matrel3}) are small 
which causes a small effect of the different signs of the matrix elements belonging to singlet and 
triplet states.

\subsubsection{Results for $M=-1$ and odd $z$-parity} \label{m1m}

{\bf a) The singlet states $\nu^1(-1)^-$}

For the singlet subspace $\nu^1(-1)^-$ we present also the ground state and the first four 
excitations, i.e. $1\le \nu \le 5$. The total energies are given in table 4. Like in the case of the 
excited $^1(-1)^+$ states there exist no data for finite field strengths in the literature. Our 
field-free data are in good agreement with the literature. In figure 2 we show the one-particle 
ionization energies analogously to figure 1. The threshold $T(B)$ is identical to the case of 
$(-1)^+$, $0^+$ or $0^-$ symmetry because the energetically lowest way to realize the ionized 
system He$^+ + e^-$ with $(-1)^-$ symmetry is to leave one electron in the $0^+$ state of a He atom. 
The other electron must then be placed in a Landau orbital with $m=-1$ and negative $z$-parity, 
which does not affect the Landau energy $B/2$.
\vspace{3mm}

\noindent
{\bf b) The triplet states $\nu^3(-1)^-$}

For the triplet states, several investigations exist in the literature 
\cite{ceperleyHF,ceperleyRPQMC,ivanov94,thurner,jones99}, in contrast to the singlet case. In table 
5 we have listed our total energies and the corresponding values of the literature for the states 
$\nu^3_{-1}(-1)^-$ (i.e. $S_z = -1$), $1\le \nu \le 5$. Again our results are better than almost all 
the reference values for finite field strengths or are at least comparable to the ones of Jones {\it
et al} \cite{jones99} within their statistical error bars. For the states $3^3(-1)^+$ and 
$4^3(-1)^+$ there exist no data in the literature so far.

In figure 2 we show the ionization energies. The dashed triplet curves almost coincide with the 
corresponding singlet curves. Such a small singlet-triplet splitting is in good agreement with the 
considerations mentioned in sec.\ref{m1p} which predict a small singlet-triplet splitting for high 
excitations.
\vspace{3mm}

\section{Transitions}

For the comparison of the energy levels of helium with the spectra of magnetic white dwarfs in 
general and GD229 in particular it is necessary to determine the transition energies from our total 
energies. Restricting ourselves to electric dipole transitions, we have the selection rules 
$\Delta S = 0$, $\Delta S_z = 0$ for the spin degrees of freedom in our nonrelativistic approach and 
$\Delta M = 0, \;\Delta \Pi_z = \pm 1$ (for linearly polarized radiation) or 
$\Delta M = \pm 1, \;\Delta \Pi_z = 0$ (for circularly polarized radiation) for the spatial degrees 
of freedom. Whereas in ref.\cite{helium} we were already able to present the $\Delta M = 0$ 
transitions $0^+ \leftrightarrow 0^-$, we are now able to investigate three times as many 
transitions: firstly the $\Delta M = 0$ transitions between the $(-1)^+$ and the $(-1)^-$ states, 
and additionally two classes of $|\Delta M| = 1$ transitions between the $M=-1$ states and $M=0$ 
states, involving positive and negative $z$-parities, respectively. Altogether our data yield $75$ 
singlet and $70$ triplet transitions.

Due to the fact that the field strengths in the atmospheres of magnetic white dwarfs is not a 
constant but varies by a factor of two for a dipole geometry, transitions which behave monotonically 
as a function of the varying field are smeared out, i.e. are not expected to provide a signature in 
the observed spectrum. However, the transitions whose wavelengths are stationary with respect to the 
field dependence manifest themselves as absorption edges in the observable spectrum if they possess 
a relevant intensity. We therefore give in tables 6 to 11 a complete list of all the stationary 
points which resulted from our calculated transitions. 

We remark that for a reliable comparison with observational spectra in some cases the finite nuclear 
mass effects have to be taken into account. This requires corrections of our total energies according 
to the scaling relation given in eq.(4) in ref.\cite{helium} and in refs.\cite{ruder,pavlov}. Due the 
selection rules $\Delta S = 0$, $\Delta S_z = 0$, we have for the transition energies the scaling 
relation $\Delta E(M_0,\mu^2 B) = \mu\Delta E(\infty,B) - \frac{1}{M_0}\mu^2 B \Delta M$ (here 
$M_0=7344a.u.$ is the nuclear mass and $\mu=0.999864a.u.$ is the reduced mass). For $\Delta M=0$ 
transitions the effect is always such that the position (field strength) and wavelength of a 
mass-corrected stationary point are related to the corresponding fixed-nucleus result by 
$B(M_0) = \mu^2B(\infty)$ and $\lambda(M_0)$ = $\frac{1}{\mu}\lambda(\infty)$. If for 
$|\Delta M|=1$ transitions the ratio $\frac{B}{M_0}$ is small compared to $\Delta E$ it is still 
possible to do an approximate correction for the wavelengths of the stationary points directly with 
the data presented: We then have 
$\lambda(M_0) = \frac{1}{\mu}\lambda(\infty)(1+\mu\frac{B}{M_0}\lambda(\infty)\Delta M)$. The 
stationary points corrected exactly within the scaling relation given above can, of course, only be 
obtained by separately scaling all the values of $\Delta E$ and by interpolating them over the grid 
of scaled field strengths. In the argumentation above we have taken into account the normal mass
correction terms. The specific mass corrections are expected to be even less significant, in 
particular for stronger fields and excited states.

Altogether we detected 139 stationary points; several ones among them possess large uncertainties 
which arise mainly due to the interpolation error with respect to the crude grid of field strengths. 
Most of the transitions $(-1)^+ \leftrightarrow 0^+$, however, are so precise that the corresponding 
data have already successfully been used to explain the absorption edges in the spectrum of the 
white dwarf GD229\cite{jssbecken}, and these data together with the other stationary transitions 
serve as a good basis for astrophysicists to investigate the spectra of unidentified magnetic 
objects.
\vspace{3mm}

\section{Concluding remarks and Outlook}

We have investigated the fixed-nucleus electronic structure of the helium atom in a magnetic field 
by a fully correlated approach. Scaling laws allow to include finite-mass effects. The present work 
is concerned with the energy levels and transitions of the helium states with magnetic quantum
number $M=-1$. A small part of the corresponding data has already been used successfully for 
identifying the features in the spectrum of the white dwarf GD229 with electronic transitions in 
atomic helium\cite{jssbecken}, which represented one of the goals of our work. The enlarged data 
presented in this paper will now, together with the energy levels provided in ref.\cite{helium}, 
serve as a good starting point for the analysis of observed spectra of magnetic astrophysical 
objects in general.

The reliability of our wavelength data results from the high accuracy of our energy values 
which ranges between $10^{-4}a.u.$ and $10^{-6}a.u.$. This accuracy has become possible due to our 
appraoch by means of an optimized anisotropic Gaussian basis set. Since the spherical invariance is 
broken by the magnetic field, it has been necessary to use Gaussians with different length scales for 
the longitudinal and transversal degrees of freedom. The nonlinear parameters describing these length 
scales have been determined by the requirement to solve optimally the one-particle problem of the H 
atom or the He${}^+$ ion in a magnetic field of given strength. These optimized one-particle
functions have been used to construct configurations in order to represent the full fixed-nucleus 
Hamiltonian and the overlap as matrices separately in each subspace of fixed quantum numbers 
corresponding to the four conserved quantities: the total spin $\mbf{S}^2$ and its $z$-component 
$S_z$, the $z$-component $L_z$ of the electronic angular momentum and the electronic $z$-parity 
$\Pi_z$. The corresponding generalized eigenvalue problem provided a variational estimation for the 
energy eigenvalues. 

Atomic energies of helium have been calculated for the ground state and the first four excitations 
in each subspace for $M=-1$, i.e. for positive and negative $z$-parity as well as for singlet and 
triplet spin symmetry. We considered altogether $20$ different field strengths $0\le B \le 100a.u.$, 
i.e. $0\le B \le 2.3505 \cdot 10^7$ Tesla. This series production of data has become by very 
efficient algorithms for the computation of the matrix elements, in particular the electon-electron 
matrix elements for which we presented an analytical formula derived in cylindrical coordinates, 
thereby making the CPU effort for the calculations for nonzero angular momentum affordable: Building 
up a matrix of dimension about 4000 takes less than one hour CPU time on a moderate Silicon Graphics 
workstation.

The comparison of energy data with observed spectra of astrophysical objects is possible by 
searching for stationary points of the transitions with respect to the magnetic field strength. The 
data for $M=-1$ in the present paper yield $\Delta M=0$ transitions, and together with the results 
for $M=0$ in ref.\cite{helium} we have also been able to consider $\Delta M=1$ transitions. Complete 
tables of all the detected stationary points for the mentioned transitions were given. Energy data 
for $M=-2$ and $M=-3$ are planned to be investigated in a future work.

In order to complete the treatment of bound electronic transitions of helium in a magnetic field, we will 
in the near future also investigate in detail the oscillator strengths of the mentioned transitions as a 
function of the magnetic field. Based on the calculated intensities it will be possible to produce 
synthetic spectra, and a comparison with the observed spectra will give important hints for models 
of the radiation transport and the field configuration in magnetic astrophysical objects.
\vspace{5mm}

{\it Acknowledgements.} The Deutsche Studienstiftung (W.B.) and the Deutsche Forschungsgemeinschaft 
(W.B.) are gratefully acknowledged for financial support.

\begin{appendix}
\renewcommand{\theequation}{\Alph{section}.\arabic{equation}}
\setcounter{equation}{0}

\section{Analytical solution to the electron-electron integral}\label{appendixA}

In the following it is our aim to derive an analytical expression for the electron-electron integral 
which allows its efficient numerical implementation. We emphasize that an efficient treatment of the 
two-particle integrals is essential for the calculations on helium since the number of two-particle 
matrix elements is $N(N+1)/2$, in contrast to the one-particle matrix elements whose number is only 
$n(n+1)/2$ (here $N\approx 4000$ is the dimension of the two-particle Hamiltonian matrix whereas 
$n\approx 200$ is the dimension of the underlying one-particle basis set, see 
eqs.(\ref{onebas},\ref{twobas})). Denoting the two-particle interaction with 
$V_{\roemII}(\mbf{r}_1,\mbf{r}_2) = \frac{1}{|\mbf{r}_1-\mbf{r}_2|}$, we have to solve the integral
\begin{eqnarray}
    \left\langle ij |V_{\roemII}|kl \right\rangle 
&=& \int d^3r_1\;d^3r_2\;\;\Phi_i(\mbf{r}_1)\Phi_j(\mbf{r}_2)\;\frac{1}{|\mbf{r}_1-\mbf{r}_2|}\;
                           \Phi_k(\mbf{r}_1)\Phi_l(\mbf{r}_2) \label{elelint},
\end{eqnarray}
where the one-particle orbitals $\Phi_i$ are of type (\ref{onebas}), obeying the constraints 
(\ref{gerade},\ref{zpar}). For the sake of brevity, we will in the following use the index notation 
$\gamma_{ik} := \gamma_i + \gamma_k$ for the sum of two indexed quantities. 

The initial step is to apply a Singer transformation \cite{singer} in order to remove the Coulomb 
singularity, thereby introducing the new variable $u$ according to 
$\frac{1}{|\mbf{r}_1-\mbf{r}_2|}=\frac{2}{\sqrt{\pi}}\int_0^{\infty}du\;e^{-u^2(\mbf{r}_1-\mbf{r}_2)^2}$. 
Then the integrand of this new integration over $u$ decomposes into a transversal and a longitudinal 
part
\begin{eqnarray}
    \left\langle ij |V_{\roemII}|kl \right\rangle
&=& \frac{2}{\sqrt{\pi}} \int\limits_{0}^{\infty}du\; I_{\rho \varphi}(u) \cdot I_{z}(u) 
                                                                                  \label{IrhophiIz},
\end{eqnarray}
where
\begin{eqnarray}
	I_z(u)   
&=&   \int\limits_{-\infty}^{\infty} dz_1
      \int\limits_{-\infty}^{\infty} dz_2\;z_1^{{n_z}_{ik}} z_2^{{n_z}_{jl}}e^{-\beta_{ik} 
                                           z_1^2-u^2(z_1-z_2)^2-\beta_{jl} z_2^2}         \label{Iz}
\end{eqnarray}
and 
\begin{eqnarray}
    I_{\rho\varphi}(u)      
 &=&     \int \rho_1 d\rho_1 d\varphi_1\; \rho_2 d\rho_2 d\varphi_2 \;\;
              \rho_1^{{n_\rho}_{ik}} e^{-i(m_i-m_k)\varphi_1}e^{-\alpha_{ik}\rho_1^2}\cdot  
			e^{-u^2(\rho_1^2+\rho_2^2 - 2\rho_1\rho_2\cos(\varphi_2-\varphi_1))}      \\
 &&  \hspace{3.3cm} \times\;\;\rho_2^{{n_\rho}_{jl}}e^{-i(m_j-m_l)\varphi_2}e^{-\alpha_{jl}\rho_2^2}    
                                                                              \label{irho} \nonumber 
\end{eqnarray}
The longitudinal integral $I_z(u)$ is the trivial part of the matrix element (\ref{elelint}). For 
decoupling the particles $1$ and $2$ we subsitute $\tilde{z}_1\!=\! z_1\! -\! b(u)z_2$, 
$\;\tilde{z}_2\!=\! z_2$ where $b(u)\!=\!\frac{u^2}{\beta_{ik}+u^2}$ and 
$\frac{\partial(\tilde{z}_1,\tilde{z}_2)}{\partial(z_1,z_2)}=1$. The exponential factorizes, and the 
power $z_1^{{n_z}_{ik}}$ can be multiplied out, yielding for $I_z(u)$ a sum over standard integrals 
of the Gaussian type $\int_{-\infty}^{\infty}z^{n_z}e^{-\gamma(u)z^2}$ which can easily be 
evaluated, giving after a few steps of algebra
\begin{eqnarray}
	I_z(u)   
&=&   4 g_{{n_z}_{ijkl}} \sum\limits_{\zeta=0}^{\zeta\le \frac{{n_z}_{ik}}{2}} 
      \left( \hspace{-2mm}\begin{array}{c} {{n_z}_{ik}} \\[-1.5mm] 2\zeta \end{array}\hspace{-2mm} 
      \right)
			J({n_z}_{ik},{n_z}_{jl},\beta_{ik},\beta_{jl};2\zeta;u)          \label{I_z}
\end{eqnarray}
where the prefactor $g_l:=mod(l,2)$ reflects the fact that the total $z$-parity is a conserved 
quantity. The function $J$ is defined by 
\begin{eqnarray}
J(n_1,n_2,a_1,a_2;v;u)    
&=&   \smf{1}{4} a_1^{-n_1-\frac{1}{2}}a_2^{-\frac{n_{12}+1}{2}} 
                 \cdot \Gamma(\smf{v+1}{2}) 
                 \cdot \Gamma(\smf{n_1+n_2-v+1}{2})
                 \cdot (a_1a_2)^{\frac{v}{2}} \nonumber \\
&&     \times\;	u^{2(n_1-v)} \;(1+\smf{1}{a_1}u^2)^{\frac{n_2-n_1}{2}} 
                               (1+\smf{a_{12}}{a_1a_2}u^2)^{-\frac{n_{12}-v+1}{2}}         \label{J}
\end{eqnarray}
The usual procedure for the treatment of the transversal part $I_{\rho\varphi}$ would be to 
represent it in Cartesian coordinates since then its decomposition into a sum of products of 
integrals $I_x(u)$ and $I_y(u)$ analoguos to $I_z(u)$ would be technically simple: The exponential 
would factorize automatically, and the remaining part consists of factors like 
$\rho^{n_\rho}e^{\pm i m} $ which can be expressed as $(x^2+y^2)^{k} (x\pm i y)^{|m|}$ due to the 
contraint (\ref{gerade}) and which can trivially be multiplied out. We emphasize that this 
procedure, however, would yield expressions which are very expensive with respect to the CPU time 
since already for small nonzero maqnetic quantum numbers $m$ or excitations $k$ the number of 
binomic terms gives rise to a large number of integrals over $u$.

In the following we exploit that the described drawback is not at all a property of the integral 
(\ref{elelint}) but only of the Cartesian approach. In fact, the many mentioned $u$-integrals are 
not independent from each other but the information how to resummarize their results to a more 
compact expression is lost in the lengthy Cartesian algebra. Therefore, it is our strategy to use 
cylindrical coordinates from the very beginning, although the derivation of this condensed 
analytical formula is technically involved.

The key for solving the integral $I_{\rho\varphi}(u)$ is the following substitution 
\begin{eqnarray}
	\tilde{\rho}_1         
&=&  \sqrt{\rho_1^2 + a^2\rho_2^2 - 2a\rho_1\rho_2\cos(\varphi_2-\varphi_1)}  \hspace{5.3cm}  
                                            \tilde{\rho}_2        \;\;=\;\;  \rho_2  \label{sub1} \\
	\sin \tilde{\varphi}_1 
&=&  \frac{\rho_1}{\tilde{\rho}_1} \sin(\varphi_2-\varphi_1) \;;\;\;\; 
        \cos\tilde{\varphi}_1 
\;\;=\;\; \frac{\rho_1}{\tilde{\rho}_1} \cos(\varphi_2-\varphi_1) - a\;\frac{\rho_2}{\tilde{\rho}_1}
     \hspace{2cm}\tilde{\varphi}_2   \;\;=\;\;  \smf{1}{2}(\varphi_2+\varphi_1) \;\;\;\;\label{sub2}
\end{eqnarray}
which results from the following ideas (here $a(u) = \frac{u^2}{\alpha_{ik}+u^2}$). Firstly, we use 
the angular part of $I_{\rho \varphi}(u)$ for exploiting the conservation of $L_z$, and secondly the 
remaining radial part has to be decoupled similarly like in the case of the $z$-integration 
mentioned above. Using in eq.(\ref{irho}) the angle $\varphi := \varphi_2-\varphi_1$ and the cyclic 
angle $\bar{\varphi} := \frac{1}{2}(\varphi_2+\varphi_1)$, the integration over $\bar{\varphi}$ 
yields a Kronecker Delta reflecting the conservation of $L_z$:
\begin{eqnarray}
	I_{\rho \varphi}(u) 
&=&  2\pi\; \delta_{m_{ij},m_{lk}}
            \int\limits_0^{\infty}\rho_1 d\rho_1 
            \int\limits_0^{\infty}\rho_2 d\rho_2 \;\;\rho_1^{{n_\rho}_{ik}} e^{-\alpha_{ik}\rho_1^2}
                                                \cdot\rho_2^{{n_\rho}_{jl}} e^{-\alpha_{jl}\rho_2^2}
                                                \cdot  e^{-u^2(\rho_1^2+\rho_2^2)}      \nonumber \\
&& \hspace{2cm}\times \;\;\int\limits_0^{2\pi}d\varphi\;e^{i(m_i-m_k)\varphi}\cdot     
e^{-u^2( - 2\rho_1\rho_2\cos \varphi)}\;\;.
\end{eqnarray}
Now, the expression $(\rho_1\cos \varphi)$ plays a similar role as $z_1$ in eq.(\ref{Iz}), and 
therefore we temporarily introduce $\xi_1=(\rho_1\cos \varphi)$ and $\eta_1=(\rho_1\sin \varphi)$ as 
dummy Cartesian variables. For decoupling the particles $1$ and $2$ analogously to the treatment of 
$I_z(u)$ we substitute $\tilde{x}_1 = \xi_1-a\tilde{\rho}_2$ whereas $\tilde{y}_1 = \eta_1$ and 
$\tilde{\rho}_2 =  \rho_2$ remain unchanged. The variables $\tilde{\rho}_1$ and $\tilde{\varphi}_1$ 
given in eqs.(\ref{sub1},\ref{sub2}) are now just the new cylindrical coordinates belonging to 
$\tilde{x}_1$ and $\tilde{y}_1$. In this new representation $I_{\rho\varphi}$ reads
\begin{eqnarray}
	I_{\rho \varphi}(u) 
&=& 2\pi\; \delta_{m_{ij},m_{lk}} \int\limits_{0}^{2\pi}d\tilde{\varphi}_1 
                                  \int\limits_{0}^{\infty} d\tilde{\rho}_1
	    \tilde{\rho}_1 \cdot \rho_1^{{n_\rho}_{ik}} e^{-(\alpha_{ik}+u^2)
            \tilde{\rho}_1^2} \cdot                     e^{i(m_i-m_k)\varphi_1}         \nonumber \\
&&  \hspace{2.5cm}\times \;  \int\limits_{0}^{\infty} d\tilde{\rho}_2 \;\; 
        \tilde{\rho}_2^{{n_\rho}_{jl}+1}
e^{- \frac{\alpha_{ik}\alpha_{jl} + \alpha{ijkl}u^2}{\alpha_{ik}+u^2}\tilde{\rho}_2^2} \;\;,   
                                                                                     \label{Irhophi}
\end{eqnarray}
where the variables $\rho_1$ and $\varphi_1$ occuring in the factor 
$\rho_1^{{n_\rho}_{ik}}e^{i(m_i-m_k)\varphi_1}$ have to be considered as functions of the variables 
labelled with the symbol ($\;\tilde{}\; $). We split 
$\rho_1^{{n_\rho}_{ik}}e^{i(m_i-m_k)\varphi_1} = \rho_1^{2k_{ik}}\cdot \rho_1^{|m_i|+|m_k|}
e^{i(m_i-m_k)\varphi_1}$ according to eq.(\ref{gerade}). This is a powerful step since the relations 
(\ref{sub1},\ref{sub2}) can explicitely be solved with respect to those two factors. After some 
algebra we arrive at:
\begin{eqnarray}
	\rho_1^{2k_{ik}} &=& \sum\limits_{r_1+r_2+r_3+r_4=k_{ik}}
\left( \hspace{-2mm}\begin{array}{c} k_{ik} \\[-1.5mm] r_1\;\;r_2\;\;r_3\;\;r_4 \end{array}\hspace{-2mm} 
\right)
				a^{2r_2+r_3+r_4}
				\tilde{\rho}_1^{2r_1+r_3+r_4}
				\tilde{\rho}_2^{2r_2+r_3+r_4}
				e^{i(r_3-r_4)\tilde{\varphi}_1}  \label{rho}
\end{eqnarray}
\begin{eqnarray}
	\rho_1^{ |m_i| + |m_k|}e^{i(m_i-m_k)\varphi_1} &=& \sum\limits_{\mu_i=0}^{|m_i|}
\left( \hspace{-2mm}\begin{array}{c}|m_i|  \\[-1.5mm] \mu_i  \end{array}\hspace{-2mm} \right)
							   \sum\limits_{\mu_k=0}^{|m_k|}
\left( \hspace{-2mm}\begin{array}{c}|m_k|  \\[-1.5mm] \mu_k  \end{array}\hspace{-2mm} \right)
		a^{|m_i|+|m_k|-\mu_{ik}}\; \tilde{\rho}_1^{\mu_{ik}}  
		\tilde{\rho}_2^{|m_i|+|m_k|-\mu_{ik}}
		\; e^{i(s_i\mu_i-s_k\mu_k)\tilde{\varphi}_1}\;\;\;\;\;                \label{rhophi}
\end{eqnarray}
where $s_i := sgn(m_i)$ are the signs of the magnetic quantum numbers. Inserting the expressions 
(\ref{rho},\ref{rhophi}) into eq.(\ref{Irhophi}), we can perform the $\tilde{\varphi}_1$-integration 
yielding a Kronecker Delta which restricts the summation indices to $r_3-r_4 = -s_i\mu_i+s_k\mu_k$. 
The obtained sum is a decomposition of $I_{\rho\varphi}$ into products of each two radial Gaussian 
integrals completely analogous to the expressions encountered in the evaluation of the integral 
$I_z$, which allows us after several steps to express also $I_{\rho\varphi}$ in terms of the 
function $J$ given in eq.(\ref{J}):
\begin{eqnarray}
	I_{\rho\varphi}(u)  &=&  4\pi^2 \,\,\delta_{m_{kl},m_{ij}}
	                \sum\limits_{\mu_i=0}^{|m_i|}
\left( \hspace{-2mm}\begin{array}{c}|m_i|  \\[-1.5mm] \mu_i  \end{array}\hspace{-2mm} \right)
			\sum\limits_{\mu_k=0}^{|m_k|}\left( \hspace{-2mm}
\begin{array}{c}|m_k|  \\[-1.5mm] \mu_k  \end{array}\hspace{-2mm} \right)
			\sum\limits_{r_1+r_2+r_3+r_4=k_{ik} \atop r_3-r_4 = -s_i\mu_i+s_k\mu_k}
			\left( \hspace{-2mm}\begin{array}{c} k_{ik} \\[-1.5mm] r_1\;\;r_2\;\;r_3\;\;r_4 
\end{array}\hspace{-2mm} \right) \nonumber \\
&&   \hspace{2cm}\times\; 
J({n_\rho}_{ik}\hspace{-0.5mm}+\hspace{-0.5mm}1,\;{n_\rho}_{jl}\hspace{-0.5mm}+\hspace{-0.5mm}1,\;
\alpha_{ik},\;\alpha_{jl};\;\,\mu_{ik}\hspace{-0.5mm}+\hspace{-0.5mm}1\hspace{-0.5mm}+
\hspace{-0.5mm}2r_1\hspace{-0.5mm}+\hspace{-0.5mm}r_3\hspace{-0.5mm}+\hspace{-0.5mm}r_4;\;u) \label{I_xy}
\end{eqnarray}
The final step is to evaluate the $u$-integration after inserting eqs.(\ref{I_xy},\ref{I_z}) into 
eq.(\ref{IrhophiIz}). Considering 
\begin{eqnarray}
	\left\langle ij |V_{\roemII}|kl \right\rangle  
&=&   32\pi^{\frac{3}{2}} \delta_{m_{kl},m_{ij}} g_{{n_z}_{ijkl}}
	                \sum\limits_{\mu_i=0}^{|m_i|}
\left( \hspace{-2mm}\begin{array}{c}|m_i|  \\[-1.5mm] \mu_i  \end{array}\hspace{-2mm} \right)
			\sum\limits_{\mu_k=0}^{|m_k|}
\left( \hspace{-2mm}\begin{array}{c}|m_k|  \\[-1.5mm] \mu_k  \end{array}\hspace{-2mm} \right)
			\sum\limits_{r_1+r_2+r_3+r_4=k_{ik} \atop r_3-r_4 = -s_i\mu_i+s_k\mu_k}
\left( \hspace{-2mm}\begin{array}{c} k_{ik} \\[-1.5mm] r_1\;\;r_2\;\;r_3\;\;r_4 \end{array}\hspace{-2mm} 
\right) 
 		        \sum\limits_{\zeta=0}^{\zeta\le \frac{{n_z}_{ik}}{2}} 
\left( \hspace{-2mm}\begin{array}{c} {{n_z}_{ik}} \\[-1.5mm] 2\zeta \end{array}\hspace{-2mm} \right) \nonumber \\
&& \hspace{-2cm}\times\;\;\;  \int\limits_{0}^{\infty}du\; 
J({n_\rho}_{ik}\hspace{-0.5mm}+\hspace{-0.5mm}1,\;{n_\rho}_{jl}\hspace{-0.5mm}+\hspace{-0.5mm}1,\;
\alpha_{ik},\;\alpha_{jl};\;\,
\mu_{ik}\hspace{-0.5mm}+\hspace{-0.5mm}1\hspace{-0.5mm}+\hspace{-0.5mm}2r_1
\hspace{-0.5mm}+\hspace{-0.5mm}r_3\hspace{-0.5mm}+\hspace{-0.5mm}r_4;\;u)\cdot
                               J({n_z}_{ik},{n_z}_{jl},\beta_{ik},\beta_{jl};2\zeta;u) 	  \label{I_K} 
\end{eqnarray}
shows that the solution of the electron-electron integral is now reduced to the integration
\begin{eqnarray}
	K(n_1,n_2,a_1,a_2;v;\;\;m_1,m_2,b_1,b_2;w) 
&=& \int\limits_{0}^{\infty}du\;J(n_1,n_2,a_1,a_2;v;u)\cdot J(m_1,m_2,b_1,b_2;w;u) \;\;.  \label{Kint}
\end{eqnarray}
Up to a constant prefactor, the integrand in eq.(\ref{Kint}) is of the type 
$g(u)=u^{2n_u}\cdot(1+au^2)^{r_a}(1+bu^2)^{r_b}(1+cu^2)^{r_c}(1+du^2)^{r_d}$ where 
$a = \frac{1}{a_1}$, $b = \frac{1}{b_1}$, $c = \frac{a_{12}}{a_1a_2}$, $d = \frac{ b_{12}}{b_1b_2}$ 
are fixed real numbers given by the basis functions. The exponents depend on the summation indices 
which enter into the functions $J$, and $n_u := n_1+m_1-v-w$ is running over positive integer 
values. Further $r_a := \frac{n_2-n_1}{2}$ and $r_b := \frac{m_2-m_1}{2}$ are integers whereas 
$r_c := \frac{v-1-n_{12}}{2}$ is integer or half-integer and $r_d := \frac{w-1-m_{12}}{2}$ is always 
half-integer. The substitution $x := \frac{u^2}{1/d+u^2}$ with 
$du = \frac{1}{2} d^{-1/2}x^{-1/2}(1-x)^{-3/2}\;dx$ leads to
\begin{eqnarray}
	\int\limits_0^\infty g(u) \; du &=& \mbox{$\frac{1}{2}$} d^{-n_u-\frac{1}{2}} 
\int\limits_0^1 
x^{n_u-\frac{1}{2}}(1-x)^{-r_{abcd}-n_u-\frac{3}{2}}(1+q_ax)^{r_a}(1+q_bx)^{r_b}(1+q_cx)^{r_c}\; dx 
\label{gsubst}
\end{eqnarray}
where $r_{abcd}:= r_a+r_b+r_c+r_d$ and 
$q_a:=\frac{a}{d}-1$, $q_b:=\frac{b}{d}-1$, $q_c:=\frac{c}{d}-1$. In order to reduce this integral 
to hypergeometric functions it is now necessary to multiply out one of the three last factors. This 
is possible since the basis functions can always be interchanged in such a way that one of the 
exponents $r_a$, $r_b$ is positive, say $r_a$. Then eq.(3.211) in ref.\cite{grad} can be used to 
obtain 
\begin{eqnarray}
		\int\limits_0^\infty g(u) \; du &=&\frac{1}{2} d^{-n_u-\frac{1}{2}} 
\sum\limits_{s=0}^{r_a}
\left( \hspace{-2mm}\begin{array}{c} r_a \\[-1.5mm] s \end{array}\hspace{-2mm} \right) 
q_a^s \;\;\times \;\;\;B(-r_{abcd}-n_u-\smf{1}{2},n_u+s+\smf{1}{2}) \nonumber \\
 && \hspace{3.75cm}\times \;\;\;F_1(n_u+s+\smf{1}{2},-r_b,-r_c,s-r_{abcd};-q_b,-q_c) \label{gu}
\end{eqnarray}
where $B$ ist the beta function and $F_1$ is the Appell hypergeometric function\cite{appell}. It is 
defined as double series, 
$F_1(a,b,b';c;x,y) := \sum_{\nu=0}^{\infty}\frac{(a,\nu)(b,\nu)}{(c,\nu)(1,\nu)}\; 
{}_2F_1(a+\nu,b',c+\nu;y)\;x^{\nu}$ where 
${}_2F_1(a,b,c;z):= \sum_{\mu=0}^{\infty}\frac{(a,\mu)(b,\mu)}{(c,\mu)(1,\mu)}\,z^\mu$ is the 
Gaussian hypergeometric function and $(a,\nu):=\frac{\Gamma(a+\nu)}{\Gamma(a)}$ is the Pochhammer 
symbol. Including all prefactors, we obtain for the integral $K$:
\begin{eqnarray}
	K(n_1,n_2,a_1,a_2,v;m_1,m_2,b_1,b_2,w)  
&=& \smf{1}{16}\; a_1^{-n_1-\frac{1}{2}}a_2^{-\frac{n_{12}+1}{2}} \cdot \Gamma(\smf{v+1}{2})
    \cdot \Gamma(\smf{n_{12}-v+1}{2})\cdot (a_1a_2)^{\frac{v}{2}}  \label{K}  \\
&&  \;\times\;  b_1^{-m_1-\frac{1}{2}}b_2^{-\frac{m_{12}+1}{2}} \cdot \Gamma(\smf{w+1}{2}) 
    \cdot \Gamma(\smf{m_{12}-w+1}{2})\cdot (b_1b_2)^{\frac{w}{2}}\times \int\limits_0^\infty g(u) 
\; du  \nonumber 
\end{eqnarray}
where now the integral stands for the expression (\ref{gu}).

The final result is a {\it real} value for $\left\langle ij |V_{\roemII}|kl \right\rangle$ obtained 
by inserting eq.(\ref{K}) into eq.(\ref{I_K}):
\begin{eqnarray}
\left\langle ij |V_{\roemII}|kl \right\rangle   
&=&  \pi^{\frac{3}{2}} \delta_{m_{kl},m_{ij}} g_{{n_z}_{ijkl}}
       \cdot \alpha_{ik}^{-n_1-\frac{1}{2}}\alpha_{jl}^{-\frac{n_{12}+1}{2}}
       \cdot \beta_{ik}^{-m_1-\frac{1}{2}}\beta_{jl}^{-\frac{m_{12}+1}{2}}
       \sum\limits_{\mu_i=0}^{|m_i|}
       \left( \hspace{-2mm}\begin{array}{c}|m_i|  \\[-1.5mm] \mu_i  \end{array}\hspace{-2mm} \right)
       \sum\limits_{\mu_k=0}^{|m_k|}
       \left( \hspace{-2mm}\begin{array}{c}|m_k|  \\[-1.5mm] \mu_k  \end{array}\hspace{-2mm} \right) 
                                                                                        \nonumber \\
&&     \hspace{4mm}\times\;\hspace{-4mm}
       \sum\limits_{r_1+r_2+r_3+r_4=k_{ik} \atop r_3-r_4 = -s_i\mu_i+s_k\mu_k} \hspace{-2mm}
       \left( \hspace{-2mm}\begin{array}{c} k_{ik} \\[-1.5mm] r_1\;\;r_2\;\;r_3\;\;r_4 
                           \end{array}\hspace{-2mm} \right) 
       \Gamma(\smf{v+1}{2})\cdot \Gamma(\smf{n_{12}-v+1}{2})
       \cdot (\alpha_{ik}\alpha_{jl})^{\frac{v}{2}} \nonumber \\
&&     \hspace{4mm}\times\;\hspace{3mm}
       \sum\limits_{\zeta=0\atop w\;\;gerade}^{\zeta \le \frac{{n_z}_{ik}}{2}}\hspace{4mm} 
       \left( \hspace{-2mm}\begin{array}{c} {{n_z}_{ik}} \\[-1.5mm] 2\zeta 
                           \end{array}\hspace{-2mm} \right)
       \hspace{4mm} \Gamma(\hf+\zeta)\cdot 
       \Gamma(\smf{m_{12}+1}{2}-\zeta)\cdot (\beta_{ik}\beta_{jl})^{\zeta}              \nonumber \\
&&     \times\;\; d^{-n_u-\frac{1}{2}}\; \sum\limits_{s=0}^{r_a}\;
       \left( \hspace{-2mm}\begin{array}{c} r_a \\[-1.5mm] s \end{array}\hspace{-2mm} \right)\;q_a^s 
                        \;\;\cdot\;B(-r_{abcd}-n_u-\smf{1}{2},n_u+s+\smf{1}{2}) \nonumber \\
&&     \hspace{3.6cm}\times \;\;F_1(n_u+s+\smf{1}{2},-r_b,-r_c,s-r_{abcd};-q_b,-q_c) 
                                                                                  \label{vorlenderg} 
\end{eqnarray}
An overview over the various parameters entering into eq.(\ref{vorlenderg}) is given in appendix 
\ref{appendixC} where we briefly present the algorithm for the implementation of 
eq.(\ref{vorlenderg}). The direct implementation of eq.(\ref{vorlenderg}) yields already a factor of 
5 to 10 with respect to the increase in speed compared with the common result derived in Cartesian 
coordinates in ref.\cite{helium}. However, by systematic exploitation of the symmetries the gain 
factor of CPU can further be raised to between 20 and 40.

\section{Symmetry properties of the electron-electron integral}\label{appendixB}

In the present section we list the different types of symmetries for the expressions encountered in 
appendix \ref{appendixA}.

\subsection{External symmetries}

By an external symmetry we mean a permutation of basis functions which leave
 $\left\langle ij |V_{\roemII}|kl \right\rangle$ invariant (up to complex conjugation). Such 
symmetries are firstly
 $ \left\langle ij |V_{\roemII}|kl \right\rangle = \left\langle ji |V_{\roemII}|lk \right\rangle$ 
since the particles are indistinguishable, and secondly we have 
$ \left\langle ij |V_{\roemII}|kl \right\rangle = \left\langle kl |V_{\roemII}| ij\right\rangle^*$ 
due to the hermiticity of $V_{\roemII}$, in the case of our real matrix elements even 
$ \left\langle ij |V_{\roemII}|kl \right\rangle = \left\langle kl |V_{\roemII}| ij\right\rangle$.

\subsection{Internal symmetries}

The integral $ \left\langle ij |V_{\roemII}|kl \right\rangle$ is also invariant under any 
manipulation of the basis functions which leave the integrals $I_z(u)$ and $I_{\rho\varphi}(u)$ in 
eq.(\ref{IrhophiIz}) invariant separately. This class of internal transformations is larger than the 
class of permutations of basis functions since the parameters entering in $I_z(u)$ and 
$I_{\rho\varphi}(u)$ can be varied independently. $I_z(u)$ is invariant under the simultaneous 
exchanges $i\leftrightarrow j$ and $k\leftrightarrow l$ of indices only in the parameters $\beta$ 
and $n_z$, and $I_{\rho\varphi}(u)$ stays unchanged under the analogous interchange of only the 
parameters $\alpha$, $n_{\rho}$ and $m$. Eq.(\ref{irho}) shows a further internal symmetry of 
$I_{\rho\varphi}(u)$: if the difference $m_i-m_k$ remains unchanged the parameters $m_i$ and $m_k$ 
can be varied arbitrarily as long as the constraint (\ref{gerade}) can be fulfilled for some $k_i$ 
and $k_k$, respectively (analogously for $m_j$ and $m_l$). It is an interesting fact that the 
matrix element $ \left\langle ij |V_{\roemII}|kl \right\rangle$ in a subspace with 
$M=m_i+m_j=m_k+m_l$ possesses the same numerical value as some matrix elements involved in subspaces 
corresponding to some other value of $M$!

The internal symmetries can be exploited for achieving simultaneously 
${n_{\rho}}_{jl} \ge {n_{\rho}}_{ik}$ and ${n_{z}}_{jl} \ge {n_{z}}_{ik}$, i.e. in the function 
$g(u)$ (see eqs.(\ref{Kint},\ref{gsubst})) {\it even both} parameters $r_a$ and $r_b$ can be assumed 
to be non-negative. A very important consequence of this is that due to the definition of $F_1$ (see 
paragraph below eq.(\ref{gu})) its double series reduces to a sum over a {\it finite number} of 
expressions involving the Gaussian hypergeometric function ${}_2F_1(a,b;c;z)$ which is much simpler 
to evaluate than $F_1(a,b,b';c;x,y)$.

\subsection{Further symmetries}

Assuming both parameters $r_a$ and $r_b$ to be non-negative, eq.(\ref{gu}) is also valid after 
interchanging $r_a\leftrightarrow r_b$ (and simultaneously $q_a\leftrightarrow q_b$), representing a 
new invariance transformation of $ \left\langle ij |V_{\roemII}|kl \right\rangle$.

\section{Implementation}\label{appendixC}

In the following we present the algorithm for the implementation of eq.(\ref{vorlenderg}) for our 
calculations on helium.

1. \parbox[t]{15cm}{Due the external symmetry $\left\langle ij |V_{\roemII}|kl \right\rangle = 
                    \left\langle kl |V_{\roemII}| ij\right\rangle$ it is sufficient to compute only 
                    the upper triangle of the matrix corresponding to $V_{\roemII}$.}
\vspace{2mm}

2. \parbox[t]{15cm}{If ${n_\rho}_{jl}-{n_\rho}_{ik} < 0 $, we use the external symmetry 
                    $\left\langle ij |V_{\roemII}|kl \right\rangle = 
                    \left\langle ji |V_{\roemII}| lk\right\rangle$ and interchange globally 
                    $i \leftrightarrow j$, $k \leftrightarrow l$, which achieves $r_a\ge 0$.}
\vspace{2mm}

3. \parbox[t]{15cm}{If then ${n_z}_{jl}-{n_z}_{ik} < 0 $, we use the internal symmetry of $I_z(u)$ 
                    and interchange ${n_z}_{i}\leftrightarrow {n_z}_{j}$, 
                    ${n_z}_{k}\leftrightarrow {n_z}_{l}$ as well as 
                    ${\beta}_{i}\leftrightarrow {\beta}_{j}$, 
                    ${\beta}_{k}\leftrightarrow {\beta}_{l}$, which achieves $r_b\ge 0$.}
\vspace{2mm}

4. \parbox[t]{15cm}{We introduce the following abbreviations for the fixed parameters independent on 
                    the summation indices of eq.(\ref{vorlenderg}):
   \vspace{2mm}

\begin{tabular}{|r c l|c|c |}\hline
name    & &  definition                                                           & type & property            \\  \hline
$   n_1$ &$=$&$ {n_\rho}_{ik} + 1     $                                           & integer & $1\le n_1$       \\
$n_{12}$ &$=$&$ {n_\rho}_{ik}+{n_\rho}_{jl} + 2$                                  & even    & $2\le n_{12}$    \\
$   m_1$ &$=$&$ {n_z}_{ik}            $                                           & integer & $0\le m_1$       \\
$m_{12}$ &$=$&$ {n_z}_{ik}+{n_z}_{jl} $                                           & even    & $0\le m_{12}$    \\
$ r_a$   &$=$&$ \smf{{n_\rho}_{jl}-{n_\rho}_{ik}}{2} $                            & integer &$0\le r_a$        \\
$ r_b$   &$=$&$ \smf{{n_z}_{jl}-{n_z}_{ik}}{2} $                                  & integer &$0\le r_b$        \\
$ d  $   &$=$&$ \smf{\beta_{ik}+\beta_{jl}}{\beta_{ik}\beta_{jl}}      $          & real & $ 1 < d $           \\
$-q_a$   &$=$&$ 1-\smf{\beta_{ik}\beta_{jl}}{\alpha_{ik}(\beta_{ik}+\beta_{jl})}$ & real & $-1 < q_a$          \\
$-q_b$   &$=$&$ \smf{\beta_{ik}}{\beta_{ik}+\beta_{jl}}                         $ & real & $-1 < q_b < 0     $ \\
$-q_c$   &$=$&$ 1-\smf{(\alpha_{ik}+\alpha_{jl})\beta_{ik}\beta_{jl}}{(\beta_{ik}+\beta_{jl})\alpha_{ik}\alpha_{jl}}$ & real & $-1 < q_c$ 
                                                                                                               \\ \hline
\end{tabular}
\vspace{4mm}
}

5. \parbox[t]{15cm}{If then $r_a > r_b $, we interchange $r_a\leftrightarrow  r_b$, 
                    $q_a\leftrightarrow  q_b$, thereby minimizing the number of terms of the 
                    innermost summation in eq.(\ref{vorlenderg}).}
\vspace{2mm}

6. \parbox[t]{15cm}{Applying the internal symmetry transformation of suitably shifting the values of 
                    $m_i$ and $m_k$, we replace both $m_i$ and $m_k$ by $m_i+2s$ and $m_k+2s$, 
                    respectively, where $s$ is the highest possible integer such that 
                    $|m_i+2s|\le {n_{\rho}}_i$ and $|m_k+2s|\le {n_{\rho}}_k$. This step is very 
                    useful for decreasing the number of summations over the indices 
                    $r_1,r_2,r_3,r_4$ since raising $m_i$ and $m_k$ has the consequence that $k_i$ 
                    and $k_k$ must decrease for keeping ${n_{\rho}}_i$ and ${n_{\rho}}_k$ constant.}
\vspace{2mm}

7. \parbox[t]{15cm}{We apply the formula (\ref{vorlenderg}), and the parameters occuring therein 
                    depend on the summation indices according to the following definitions:
\vspace{2mm}

\begin{tabular}{|r c l|c|r c l|}\hline
name     & &  definition                                  & type & \multicolumn{3}{|c|}{property}           \\  \hline
$\!\!v       \!\!$&$\!\!=\!$&$\!\! \mu_{ik}\hspace{-0.5mm}+\hspace{-0.5mm}1\hspace{-0.5mm}
                                     +\hspace{-0.5mm}2r_1\hspace{-0.5mm}
     	     	         +    \hspace{-0.5mm}r_3\hspace{-0.5mm}
     	     	         +    \hspace{-0.5mm}r_4     $    & integer      & $1\le$&$\hspace{-4mm} v\hspace{-4mm}$&$ \le {n_\rho}_{ik}+1   $  \\
$\!\!w       \!\!$&$\!\!=\!$&$\!\! 2\zeta            $    & even         & $0\le$&$\hspace{-4mm} w\hspace{-4mm}$&$ \le {n_z}_{ik}-g_{{n_{z}}_{ik}}   $ \\
$\!\!n_u     \!\!$&$\!\!=\!$&$\!\! n_1+m_1- v-w $         & integer      & $g_{{n_z}_{ik}}\le$&$\hspace{-4mm} n_u\hspace{-4mm} $&$\le {n_\rho}_{ik} + {n_z}_{ik}  $  \\
$\!\!r_c     \!\!$&$\!\!=\!$&$\!\! \smf{v-1-n_{12}}{2} $  & int. or half-int. & $-\smf{{n_\rho}_{ik}+{n_\rho}_{jl}}{2} - 1 \le$&$\hspace{-4mm} r_c\hspace{-4mm} $&$\le -\smf{{n_\rho}_{jl}}{2} - 1  $  \\
$\!\!r_d     \!\!$&$\!\!=\!$&$\!\! \smf{w-1-m_{12}}{2} $  & half-integer&$  -\smf{1+{n_z}_{ik}+{n_z}_{jl}}{2} \le$&$\hspace{-4mm} r_d\hspace{-4mm} $&$\le -\smf{1+g_{{n_z}_{ik}}+{n_z}_{jl}}{2} < 0                       $  \\
$\!\!r_{abcd}\!\!$&$\!\!=\!$&$\!\! \hspace{-1mm}-n_1\hspace{-1mm}-\hspace{-0.5mm}m_1\hspace{-1mm}+\smf{v+w}{2}-1      $    &int. or half-int. & $  -{n_\rho}_{ik}\hspace{-1mm}-{n_z}_{ik}\hspace{-1mm}-\hspace{-0.5mm}\smf{3}{2} \le$&$\hspace{-4mm} r_{abcd}\hspace{-4mm} $&$\le \smf{-{n_\rho}_{ik}\hspace{-1mm}-{n_z}_{ik}\hspace{-1mm}-3-g_{{n_z}_{ik}}}{2} < 0 $  \\ \hline
\end{tabular}
}
\vspace{4mm}

8. \parbox[t]{15cm}{Although in principle $F_1(a,b,b';c;x,y)$ is a double power series in the 
                    arguments $x,y$ with convergence radii $|x|<1$ and $|y|<1$ and although in our 
                    case even both arguments may happen to lie close to $1$, we do not have to care 
                    about the power series in $x$. The reason is that the corresponding series 
                    terminates after $r_b$ terms because the negative integer $-r_b$ enters as 
                    second parameter argument into $F_1$ in eq.(\ref{vorlenderg}). The sum terms in
                    $F_1$ each involve once the Gaussian hypergeometric function, and since their 
                    arguments are related, it is possible to use a continued fraction representation 
                    for the ratio $\frac{{}_2F_1(a+1,b,c+1,z)}{{}_2F_1(a,b,c,z)}$ in order to 
                    establish a recursion law which is stable over the typically $r_b<10$ 
                    recursions. Therefore, for each occurence of $F_1$, it was only necessary to 
                    evaluate a single time the Gaussian hypergeometric function ${}_2F_1$, which is      
                    not always simple but very efficiently possible by means of {\it fast analytical 
                    continuation formulas}. Several of them are given in eqs.(15.3.3-12) in 
                    ref.\cite{abramowitz}, a much larger set of such continuation formulas is 
                    presented in ref.\cite{hyper}.}
\vspace{2mm}

We point out that without such a systematic analysis of the electron-electron integral as well as 
the Appell hypergeometric function $F_1$ and in particular the Gaussian hypergeometric function 
${}_2F_1$, the computation of many excited helium states for nonzero magnetic quantum number for 
many different field strengths would not have been possible.

\vspace{2mm}

\end{appendix}

\end{document}


\begin{center}
{\LARGE\bf Tables}
\end{center}
\vspace{1cm}

\begin{table}
\begin{minipage}{9.6cm}
\caption{ Total energies $E$ of the singlet state $1^1(-1)^+$, one-electron ionization threshold $T$ 
and, if available, best energy values given in the literature, as a function of the magnetic field 
strength $B$}
\vspace{1mm}

\begin{tabular}{|l|l|l|l|l|}\hline\hline
$B$     &  $ E(1^1(-1)^+)$    &\multicolumn{1}{|c|}{literature \rule[-3mm]{0mm}{8mm}}    &  $T$       \\ \hline
        &                  &                    &                       \\
0.0000  &  $-2.123774$   &$-2.12384343087^a$  & $-2.000      $        \\
0.0008  &  $-2.124179$   &                    & $-1.999599960$        \\    
0.004   &  $-2.125732$   &                    & $-1.997999000$        \\ 
0.008   &  $-2.127580$   &                    & $-1.995995995$        \\ 
0.020   &  $-2.132539$   &                    & $-1.989975001$        \\    
0.040   &  $-2.139008$   &                    & $-1.979900008$        \\    
0.080   &  $-2.146622$   &                    & $-1.959600176$        \\
0.160   &  $-2.148448$   &                    & $-1.918402804$        \\
0.240   &  $-2.139976$   &                    & $-1.876414090$        \\  
0.400   &  $-2.105987$   &                    & $-1.790105922$        \\
0.500   &  $-2.077302$   &$-2.0750^b$         & $-1.734628064$        \\   
0.800   &  $-1.969560$   &                    & $-1.561526260$        \\
1.000   &  $-1.884875$   &$-1.8800^b$         & $-1.440989741$        \\
1.600   &  $-1.590118$   &                    & $-1.058421519$        \\
2.000   &  $-1.368986$   &$-1.3591^b$         & $-0.788842154$        \\ 
5.000   &  $ 0.619265$   &                    & $ 1.456132354$        \\
10.000  &  $ 4.500982$   &                    & $ 5.609851957$        \\
20.000  &  $13.010551$   &                    & $14.47840453 $        \\ 
50.000  &  $40.339488$   &                    & $42.45369755 $        \\ 
100.000 &  $87.671288$   &                    & $90.43945348 $        \\   \hline
\end{tabular}
\end{minipage}
\vspace{1mm}

$^a$ Drake et al. (1992) \cite{drake}

$^b$ Larsen (1979) \cite{larsen}
\vspace{10cm}

\end{table}

\begin{table}
\begin{minipage}{15.8cm}
\caption{ Total energies $E$ of the excited singlet states $\nu^1(-1)^+$ for $2\le \nu\le 5$ as a function 
of the magnetic field strength $B$ as well as field free reference values}
\vspace{1mm}

\begin{tabular}{|l|l|l|l|l|}\hline\hline
$B$            &    $ E(2^1(-1)^+)$       &     $ E(3^1(-1)^+)$         &       $ E(4^1(-1)^+)$       &    $ E(5^1(-1)^+)$                \\ \hline
               &                          &                             &                             &                                   \\                         
0.0000         &$-2.055124     $          &$-2.031253     $             &$-2.031060           $       &$-2.020002        $                \\
0.0000 (lit.)  &$-2.055146362^a$          &$-2.031255144^a$             &$-2.03106965^a       $       &$-2.020002937^a   $                \\
0.0008         &$-2.055514     $          &$-2.031638     $             &$-2.031422           $       &$-2.020356        $                \\    
0.004          &$-2.056834     $          &$-2.032905     $             &$-2.032089           $       &$-2.021046        $                \\ 
0.008          &$-2.057978     $          &$-2.033939     $             &$-2.031432           $       &$-2.020791        $                \\ 
0.020          &$-2.058622     $          &$-2.034384     $             &$-2.024218           $       &$-2.017013        $                \\    
0.040          &$-2.053994     $          &$-2.029964     $             &$-2.013028           $       &$-2.003001        $                \\    
0.080          &$-2.038112     $          &$-2.010281     $             &$-1.994622           $       &$-1.982587        $                \\
0.160          &$-2.004984     $          &$-1.964022     $             &$-1.946526           $       &$-1.937449        $                \\
0.240          &$-1.970004     $          &$-1.923549     $             &$-1.904394           $       &$-1.893803        $                \\  
0.400          &$-1.893976     $          &$-1.840324     $             &$-1.819308           $       &$-1.808757        $                \\
0.500          &$-1.843343     $          &$-1.786317     $             &$-1.764444           $       &$-1.753581        $                \\   
0.800          &$-1.680951     $          &$-1.616388     $             &$-1.592658           $       &$-1.581405        $                \\
1.000          &$-1.565692     $          &$-1.497369     $             &$-1.472743           $       &$-1.461154        $                \\
1.600          &$-1.194554     $          &$-1.117980     $             &$-1.091462           $       &$-1.079247        $                \\
2.000          &$-0.930508     $          &$-0.849891     $             &$-0.822478           $       &$-0.809928        $                \\ 
5.000          &$ 1.291512     $          &$ 1.389207     $             &$ 1.420194           $       &$ 1.433857        $                \\
10.000         &$ 5.428064     $          &$ 5.538802     $             &$ 5.572339           $       &$ 5.586918        $                \\
20.000         &$14.279839     $          &$14.403512     $             &$14.439480           $       &$14.455341        $                \\ 
50.000         &$42.233602     $          &$42.374094     $             &$42.413014           $       &$42.429203        $                \\ 
100.000        &$90.203618     $          &$90.356573     $             &$90.397606           $       &$90.416255        $                \\   \hline
\end{tabular}
\end{minipage}
\vspace{1mm}

$^a$ Drake et al. (1992) \cite{drake}
\end{table}

\begin{sidewaystable}
{\small
\begin{minipage}{18.2cm}
\caption{ Total energies $E$ of the triplet states $\nu^3_{-1}(-1)^+$ ($S_z=-1$) for $1\le \nu\le 5$ as a 
function of the magnetic field strength $B$. We have also provided the results given in the literature so 
far.}
\vspace{1mm}

\begin{tabular}{|l||l      |l                ||l               |l                    ||l               |l                     ||l             |l                     |l |}\hline\hline
\rule[-3mm]{0mm}{8mm}
        &\multicolumn{2}{|c||}{ $E(1^3(-1)^+)$}&\multicolumn{2}{|c||}{ $E(2^3(-1)^+)$}&\multicolumn{2}{|c||}{ $E(3^3(-1)^+)$}  &  $E(4^3(-1)^+)$ &  $E(5^3(-1)^+)$      \\ \hline
\multicolumn{1}{|c||}{ $B$ }
        &  this work       &     literature  &  this work       &     literature      &  this work      &     literature       &  this work      &  this work           \\ \hline
        &                  &                 &                  &                     &                 &                      &                 &                      \\
0.0000  &   $-2.133149$    & $-2.133164191^a$&  $-2.058076$     &  $-2.058081084^a$   &   $-2.032322$   & $-2.032324354^a$     &   $-2.031253$   &  $-2.020550$         \\
        &                  &                 &                  &                     &                 &                      & $-2.031255168^a$& $-2.020551187^a$     \\
0.0008  &   $-2.134349 $   &   $-2.1326^c$   &  $-2.059266$     &  $-2.0588^c$        &   $-2.033487$   &                      &   $-2.032438$   &  $-2.021661$         \\
0.004   &   $-2.139109 $   &   $-2.1374^c$   &  $-2.063816$     &  $-2.0633^c$        &   $-2.037489$   &                      &   $-2.036839$   &  $-2.025180$         \\
0.008   &   $-2.144983 $   &   $-2.1437(7)^d$&  $-2.069049$     &  $-2.0689(4)^d$     &   $-2.042110$   & $-2.0407(3)^d$       &   $-2.040765$   &  $-2.028940$         \\
0.020   &   $-2.162112 $   &                 &  $-2.082183$     &                     &   $-2.054609$   &                      &   $-2.045891$   &  $-2.037572$         \\
0.040   &   $-2.189128 $   &   $-2.187300^f$ &  $-2.098532$     &  $-2.096752^f$      &   $-2.070637$   &                      &   $-2.053178$   &  $-2.044517$         \\
0.080   &   $-2.238504 $   &   $-2.2384(3)^g$&  $-2.123337$     &  $-2.1231(5)^g$     &   $-2.093337$   & $-2.0898(3)^g$       &   $-2.074828$   &  $-2.062590$         \\
0.160   &   $-2.325189 $   &                 &  $-2.169263$     &                     &   $-2.127206$   &                      &   $-2.109343$   &  $-2.099747$         \\
0.240   &   $-2.402393 $   &                 &  $-2.214084$     &  $-2.210304^f$      &   $-2.165492$   &                      &   $-2.145502$   &  $-2.134821$         \\
0.400   &   $-2.540763 $   &   $-2.5366^c$   &  $-2.298465$     &  $-2.2972^h$        &   $-2.241932$   & $-2.2413^h$          &   $-2.220053$   &  $-2.209244$         \\
0.500   &   $-2.620021 $   &   $-2.6185^b$   &  $-2.348197$     &                     &   $-2.287934$   &                      &   $-2.265163$   &  $-2.254042$         \\ 
0.800   &   $-2.835619 $   &   $-2.8356(2)^g$&  $-2.486853$     &  $-2.4869(11)^g$    &   $-2.418154$   & $-2.4189(2)^g$       &   $-2.393403$   &  $-2.381809$         \\
1.000   &   $-2.965504 $   &   $-2.9638^b$   &  $-2.572178$     &                     &   $-2.499237$   &                      &   $-2.473517$   &  $-2.461576$         \\  
1.600   &   $-3.308774 $   &   $-3.3079(7)^d$&  $-2.802301$     &  $-2.8015(8)^d$     &   $-2.720071$   & $-2.6950(12)^d$      &   $-2.692304$   &  $-2.679694$         \\
2.000   &   $-3.508911 $   &   $-3.5063^b$   &  $-2.938822$     &                     &   $-2.852078$   &                      &   $-2.823349$   &  $-2.810394$         \\ 
5.000   &   $-4.625491 $   &   $-4.6173^e$   &  $-3.718569$     &                     &   $-3.613233$   &                      &   $-3.580744$   &  $-3.566622$         \\ 
10.000  &   $-5.839475 $   &   $-5.8295^e$   &  $-4.582689$     &                     &   $-4.463680$   &                      &   $-4.428595$   &  $-4.413579$         \\
20.000  &   $-7.440556 $   &   $-7.4277^e$   &  $-5.731115$     &                     &   $-5.598915$   &                      &   $-5.561423$   &  $-5.545260$         \\
50.000  &   $-10.28410 $   &   $-10.2644^e$  &  $-7.777100$     &                     &   $-7.628163$   &                      &   $-7.587802$   &  $-7.571207$         \\
100.00  &   $-13.10478 $   &   $-13.0764^e$  &  $-9.806638$     &                     &   $-9.645518$   &                      &   $-9.603141$   &  $-9.584339$         \\ \hline
\end{tabular}

\end{minipage}
\vspace{1mm}

$^a$ Drake et al. (1992) \cite{drake}

$^b$ Larsen (1979) \cite{larsen}

$^c$ Jones et al. (1996) \cite{ceperleyHF}

$^d$ Jones et al. (1997) \cite{ceperleyRPQMC}

$^e$ Ivanov (1994) \cite{ivanov94}

$^f$ Thurner et al. (1993) \cite{thurner}

$^g$ Jones et al. (1999) \cite{jones99}

$^h$ ETG Hartree-Fock energies (Jones et al. (1999) \cite{jones99})
}

\end{sidewaystable}

\begin{sidewaystable}
{\small
\begin{minipage}{18.5cm}
\caption{ Total energies $E$ of the states $\nu^1(-1)^-$, $1\le \nu \le 5$ as a function of the magnetic 
field strength $B$ as well as field free reference values. Reference values for finite field strength are
not available in the literature so far.}
\vspace{1mm}

\begin{tabular}{|l|l|l|l|l|l|}\hline\hline
$B$     &$ E(1^1(-1)^-)$   & $ E(2^1(-1)^-)$ &$ E(3^1(-1)^-)$ & $ E(4^1(-1)^-)$   & $ E(5^1(-1)^-) $     \\ \hline
        &                  &                 &                &                   &                      \\
0.0000  &  $-2.055613$     &$-2.031277     $ &$-2.020014  $   &$-2.020000    $    &$-2.013892   $        \\
        &$-2.055620733^a$  &$-2.031279846^a$ &$-2.020015836^a$&$-2.020000711^a$   &$-2.013898227^a$      \\
0.0008  &  $-2.056007$     &$-2.031653     $ &$-2.020375  $   &$-2.020344    $    &$-2.014196     $      \\    
0.004   &  $-2.057469$     &$-2.032717     $ &$-2.021323  $   &$-2.020478    $    &$-2.014483     $      \\ 
0.008   &  $-2.059049$     &$-2.033191     $ &$-2.021527  $   &$-2.018934    $    &$-2.013542     $      \\ 
0.020   &  $-2.062299$     &$-2.031471     $ &$-2.018372  $   &$-2.012348    $    &$-2.007346     $      \\    
0.040   &  $-2.064207$     &$-2.025446     $ &$-2.008719  $   &$-2.000788    $    &$-1.996278     $      \\    
0.080   &  $-2.060968$     &$-2.010768     $ &$-1.989904  $   &$-1.979451    $    &$-1.973516     $      \\
0.160   &  $-2.042173$     &$-1.976616     $ &$-1.951446  $   &$-1.939499    $    &$-1.932936     $      \\
0.240   &  $-2.015792$     &$-1.939171     $ &$-1.911235  $   &$-1.898364    $    &$-1.891395     $      \\  
0.400   &  $-1.951742$     &$-1.858850     $ &$-1.827207  $   &$-1.813144    $    &$-1.805743     $      \\
0.500   &  $-1.906870$     &$-1.806041     $ &$-1.772722  $   &$-1.758134    $    &$-1.750501     $      \\   
0.800   &  $-1.757739$     &$-1.638598     $ &$-1.601676  $   &$-1.585991    $    &$-1.577931     $      \\
1.000   &  $-1.649285$     &$-1.520740     $ &$-1.482091  $   &$-1.465894    $    &$-1.457643     $      \\
1.600   &  $-1.293330$     &$-1.143718     $ &$-1.101454  $   &$-1.084209    $    &$-1.075561     $      \\
2.000   &  $-1.036760$     &$-0.876696     $ &$-0.832747  $   &$-0.815025    $    &$-0.806167     $      \\ 
5.000   &  $ 1.154910$     &$ 1.358659     $ &$ 1.409032  $   &$ 1.428522    $    &$ 1.438083     $      \\
10.000  &  $ 5.271756$     &$ 5.506378     $ &$ 5.560823  $   &$ 5.581393    $    &$ 5.591327     $      \\
20.000  &  $14.109153$     &$14.370201     $ &$14.427890  $   &$14.449299    $    &$14.459536     $      \\ 
50.000  &  $42.053685$     &$42.341109     $ &$42.401827  $   &$42.424005    $    &$42.434562     $      \\ 
100.000 &  $90.023511$     &$90.324738     $ &$90.386944  $   &$90.409492    $    &$90.420161     $      \\   \hline
\end{tabular}		
\end{minipage}

\vspace{1mm}

$^a$ Drake et al. (1992) \cite{drake}

}
\end{sidewaystable}

\begin{sidewaystable}
{\small
\begin{minipage}{23.7cm}
\caption{ Total energies $E$ of the triplet states $\nu^3_{-1}(-1)^-$ ($S_z=-1$) for $1\le \nu\le 5$ as a 
function of the magnetic field strength $B$. We have also provided the results given in the literature so 
far.}
\vspace{1mm}

\begin{tabular}{|l||l      |l      ||l          |l             ||l        |l           ||l            ||l       |l        ||l     ||    }\hline\hline
\rule[-3mm]{0mm}{8mm}
      & \multicolumn{2}{|c||}{ $E(1^3(-1)^-)$} 
      & \multicolumn{2}{c||}{ $E(2^3(-1)^-)$}   
      & \multicolumn{2}{c||}{ $E(3^3(-1)^-)$}   
      & { $E(4^3(-1)^-)$}   
      & { $E(5^3(-1)^-)$}       \\ \hline
\multicolumn{1}{|c||}{ $B$ }
      & this work  &   literature    & this work &   literature    & this work &   literature     & this work &        this work          \\ \hline
      &            &                 &           &                 &           &                  &           &                           \\
0.0000& $-2.055629$& $-2.055636309^a$&$-2.031286$& $-2.031288847^a$&$-2.020020$& $-2.020021027^a$ &$-2.020000$&       $-2.013895$         \\
      &            &                 &           &                 &           &                  &$-2.020000711^a$  &$-2.013901415^a$    \\
0.0008&$-2.056824 $& $-2.056766^f$   &$-2.032463$&  $-2.032436^f$  &$-2.021176$&                  &$-2.021148$&       $-2.014998$         \\
0.004 &$-2.061486 $& $-2.061428^f$   &$-2.036727$&  $-2.036694^f$  &$-2.025324$&                  &$-2.024483$&       $-2.018484$         \\
0.008 &$-2.067062 $& $-2.067000^f$   &$-2.041202$&  $-2.0413(9)^d$ &$-2.029529$& $-2.0266(7)^d$   &$-2.026940$&       $-2.021545$         \\
0.020 &$-2.082319 $&                 &$-2.051485$&                 &$-2.038378$&                  &$-2.032348$&       $-2.027352$         \\
0.040 &$-2.104234 $&   $-2.103728^f$ &$-2.065465$&  $-2.057738^f$  &$-2.048734$&                  &$-2.040795$&       $-2.036278$         \\
0.080 &$-2.141017 $&   $-2.1414(3)^g$&$-2.090793$&  $-2.0912(2)^g$ &$-2.069919$& $-2.0696(4)^g$   &$-2.059460$&       $-2.053523$         \\
0.160 &$-2.202291 $&   $-2.193460^f$ &$-2.136663$&                 &$-2.111468$&                  &$-2.099511$&       $-2.092944$         \\
0.240 &$-2.256006 $&   $-2.238396^f$ &$-2.179246$&                 &$-2.151267$&                  &$-2.138381$&       $-2.131405$         \\
0.400 &$-2.352208 $&   $-2.3512^h$   &$-2.258990$&  $-2.2588^h$    &$-2.227265$& $-2.2271^h$      &$-2.213172$&       $-2.205759$         \\
0.500 &$-2.407521 $&                 &$-2.306225$&                 &$-2.272795$&                  &$-2.258169$&       $-2.250521$         \\ 
0.800 &$-2.559005 $&   $-2.5577(4)^g$&$-2.438913$&  $-2.4392(5)^g$ &$-2.401794$& $-2.4022(4)^g$   &$-2.386047$&       $-2.377963$         \\
1.000 &$-2.650973 $&                 &$-2.521134$&                 &$-2.482236$&                  &$-2.465962$&       $-2.457680$         \\  
1.600 &$-2.896192 $&   $-2.8964(7)^d$&$-2.744310$&  $-2.7394(8)^d$ &$-2.701661$& $-2.6436(17)^d$  &$-2.684304$&       $-2.675612$         \\
2.000 &$-3.040304 $&                 &$-2.877390$&                 &$-2.832986$&                  &$-2.815133$&       $-2.806227$         \\ 
5.000 &$-3.851883 $&                 &$-3.642446$&                 &$-3.591326$&                  &$-3.571636$&       $-3.562005$         \\ 
10.000&$-4.737490 $&                 &$-4.494996$&                 &$-4.439609$&                  &$-4.418795$&       $-4.408774$         \\
20.000&$-5.902110 $&                 &$-5.631390$&                 &$-5.572603$&                  &$-5.550913$&       $-5.540577$         \\
50.000&$-7.959094 $&                 &$-7.660647$&                 &$-7.598710$&                  &$-7.576226$&       $-7.565565$         \\
100.00&$-9.989376 $&                 &$-9.677020$&                 &$-9.613592$&                  &$-9.590738$&       $-9.579962$         \\ \hline
\end{tabular}													       

\end{minipage}
\vspace{1mm}

$^a$ Drake et al. (1992) \cite{drake}

$^d$ Jones et al. (1997) \cite{ceperleyRPQMC}

$^f$ Thurner et al. (1993) \cite{thurner}

$^g$ Jones et al. (1999) \cite{jones99}

$^h$ ETG Hartree-Fock energies (Jones et al. (1999) \cite{jones99})
}

\end{sidewaystable}

\begin{table}
\begin{minipage}{11.0cm}
\caption{Overview over all of the stationary points found in the singlet $\Delta M = 0$ transitions 
$\mu^1(-1)^+ \rightarrow \nu^1(-1)^-$ for $\mu = 1,...,5,\;\nu = 1,...,5$ in the range $0\le B \le 100a.u.$. 
The main part of the errors arises due to the interpolation over the relatively crude grid of field 
strengths. For high wavelengths, the finite accuracy of the energy values themselves contributes also to
error in the energies. }
\vspace{1mm}
\label{statsing3}
\begin{tabular}{l   l   l   l}\hline\hline
           Component                        & wavelength/A         & position $B/a.u.$    & max/min  \\ 
$\nu^{2S+1}M^{(-1)^{\Pi_z}} $               &                      &                      &          \\ \hline
$2^1(-1)^-  \rightarrow  2^1(-1)^+$         & $15915 \pm 50$       & $0.035 \pm 0.003$    & min      \\	 
$2^1(-1)^-  \rightarrow  2^1(-1)^+$         & $16770 \pm 85$       & $0.098 \pm 0.007$    & max      \\	 
$2^1(-1)^-  \rightarrow  3^1(-1)^+$         & $13649.7 \pm 3.0$    & $26.06 \pm 0.04$     & min      \\	 
$3^1(-1)^-  \rightarrow  3^1(-1)^+$         & $20505   \pm 80$     & $0.0542 \pm 0.0004$  & min      \\	 
$3^1(-1)^-  \rightarrow  3^1(-1)^+$         & $37139   \pm 20$     & $0.213 \pm 0.001$    & max      \\	 
$3^1(-1)^-  \rightarrow  4^1(-1)^+$         & $39258 \pm 50$       & $16.5 \pm 0.3$       & min      \\	 
$4^1(-1)^-  \rightarrow  4^1(-1)^+$         & $29340 \pm 550$      & $0.08  \pm  0.01$    & min      \\	 
$4^1(-1)^-  \rightarrow  4^1(-1)^+$         & $76600 \pm 600$      & $0.29  \pm  0.01$    & max      \\	 
$5^1(-1)^-  \rightarrow  4^1(-1)^+$         & $25417.4 \pm 30$     & $0.00675 \pm 0.00004$& min      \\	 
$5^1(-1)^-  \rightarrow  4^1(-1)^+$         & $27502 \pm 30$       & $0.03076 \pm 0.00001$& max      \\	 
$5^1(-1)^-  \rightarrow  4^1(-1)^+$         & $18500 \pm 1200$     & $0.059 \pm 0.002$    & min      \\
$5^1(-1)^-  \rightarrow  4^1(-1)^+$         & $35220 \pm 300$      & $0.23 \pm 0.01$      & max      \\
$5^1(-1)^-  \rightarrow  5^1(-1)^+$         & $72600 \pm  5200$    & $0.046 \pm 0.006$    & max      \\
$5^1(-1)^-  \rightarrow  5^1(-1)^+$         & $45100 \pm  4000$    & $0.095 \pm 0.006$    & min      \\
$5^1(-1)^-  \rightarrow  5^1(-1)^+$         & $208000\pm 12000$    & $0.290 \pm 0.015$    & max      \\	 
\end{tabular}
\end{minipage}
\end{table}

\begin{table}
\begin{minipage}{11.0cm}
\caption{Same as Table \ref{statsing3} but for the triplet $\Delta M = 0$ transitions $\mu^3(-1)^+ \rightarrow \nu^3(-1)^-$ 
for $\mu = 1,...,5,\;\nu = 1,...,5$ }
\vspace{1mm}
\begin{tabular}{l   l   l   l}\hline\hline
           Component                        & wavelength/A        & position $B/a.u.$   &   max/min     \\ 
        $\nu^{2S+1}M^{(-1)^{\Pi_z}}$        &                     &                     &               \\ \hline
$2^3(-1)^-  \rightarrow  3^3(-1)^+$         & $603000 \pm 11000$  & $0.00320\pm 0.00001$& max           \\	 
$2^3(-1)^-  \rightarrow  3^3(-1)^+$         & $85000  \pm  3200$  & $0.036 \pm 0.002$   & min           \\	 
$2^3(-1)^-  \rightarrow  3^3(-1)^+$         & $139400 \pm 700$    & $31 \pm 5$          & min           \\	 
$3^3(-1)^-  \rightarrow  3^3(-1)^+$         & $37550  \pm  160$   & $0.0034 \pm 0.0009$ & max           \\	 
$3^3(-1)^-  \rightarrow  3^3(-1)^+$         & $17700  \pm 300$    & $0.059 \pm 0.001$   & min           \\	 
$3^3(-1)^-  \rightarrow  3^3(-1)^+$         & $32200  \pm 60$     & $0.270 \pm 0.006$   & max           \\	 
$3^3(-1)^-  \rightarrow  4^3(-1)^+$         & $39240  \pm 200$    & $0.0045 \pm 0.0015$ & min           \\	 
$3^3(-1)^-  \rightarrow  4^3(-1)^+$         & $40751  \pm 10$     & $20.0 \pm 0.5$      & min           \\	 
$4^3(-1)^-  \rightarrow  3^3(-1)^+$         & $12930  \pm 240 $   & $0.063 \pm 0.003$   & min           \\
$4^3(-1)^-  \rightarrow  3^3(-1)^+$         & $16864  \pm  33$    & $0.215 \pm 0.004$   & max           \\
$4^3(-1)^-  \rightarrow  4^3(-1)^+$         & $32340  \pm  270$   & $0.012 \pm 0.0006$  & min           \\
$4^3(-1)^-  \rightarrow  4^3(-1)^+$         & $38800  \pm  1600$  & $0.03 \pm 0.002$    & max           \\
$4^3(-1)^-  \rightarrow  4^3(-1)^+$         & $29460  \pm  200$   & $0.081 \pm 0.008$   & min           \\
$4^3(-1)^-  \rightarrow  4^3(-1)^+$         & $66600  \pm  200$   & $0.34 \pm 0.01$     & max           \\	 
$4^3(-1)^-  \rightarrow  5^3(-1)^+$         & $87175  \pm 60$     & $0.0196 \pm 0.0002$ & min           \\
$5^3(-1)^-  \rightarrow  3^3(-1)^+$         & $11170  \pm 180$    & $0.065 \pm 0.004$   & min           \\
$5^3(-1)^-  \rightarrow  3^3(-1)^+$         & $13470  \pm 30$     & $0.2002 \pm 0.0009$ & max           \\
$5^3(-1)^-  \rightarrow  4^3(-1)^+$         & $23630  \pm 40$     & $0.0098 \pm 0.0005$ & min           \\
$5^3(-1)^-  \rightarrow  4^3(-1)^+$         & $28200  \pm 1000$   & $0.031 \pm 0.003$   & max           \\
$5^3(-1)^-  \rightarrow  4^3(-1)^+$         & $21050  \pm 520$    & $0.088 \pm 0.01$    & min           \\
$5^3(-1)^-  \rightarrow  4^3(-1)^+$         & $32680  \pm 340$    & $0.23 \pm 0.03$     & max           \\
$5^3(-1)^-  \rightarrow  5^3(-1)^+$         & $42200  \pm 1700$   & $0.015 \pm 0.001$   & min           \\
$5^3(-1)^-  \rightarrow  5^3(-1)^+$         & $55790  \pm  320$   & $0.046 \pm 0.002$   & max           \\
$5^3(-1)^-  \rightarrow  5^3(-1)^+$         & $48400  \pm  960$   & $0.100 \pm 0.004$   & min           \\
$5^3(-1)^-  \rightarrow  5^3(-1)^+$         & $144000 \pm  8000$  & $0.30 \pm 0.01$     & max           \\
$5^3(-1)^-  \rightarrow  5^3(-1)^+$         & $80200  \pm   500$  & $49.4 \pm 7$        & min           \\
\end{tabular}
\end{minipage}
\end{table}

\begin{table}
\begin{minipage}{11.0cm}
\caption{Overview over all of the stationary points found in the singlet $|\Delta M| = 1$ transitions
within the subspace of positive $z$-parity, 
$\mu^1(-1)^+ \rightarrow \nu^10^+$ for $\mu = 1,...,5,\;\nu = 1,...,5$ in the range $0\le B \le 100a.u.$.}
\vspace{1mm}
\label{statsing}
\begin{tabular}{l   l   l   l}\hline\hline
           Component                        & wavelength/A         & position $B/a.u.$      & max/min  \\ 
$\nu^{2S+1}M^{(-1)^{\Pi_z}} $               &                      &                        &          \\ \hline
$ 1^10^+ \rightarrow 1^1(-1)^+ $            &$ 608.1  \pm 0.3  $   &  $ 0.1672 \pm 0.0015 $ & max      \\	 
$ 1^10^+ \rightarrow 2^1(-1)^+ $            &$ 539.95 \pm 0.25 $   &  $ 0.0160 \pm 0.0007 $ & max      \\	 
$ 1^10^+ \rightarrow 3^1(-1)^+ $            &$ 524.83 \pm 0.24 $   &  $ 0.017  \pm 0.002  $ & max      \\	 
$ 1^10^+ \rightarrow 4^1(-1)^+ $            &$ 523.37 \pm 0.22 $   &  $ 0.004  \pm 0.001  $ & max      \\	 
$ 1^10^+ \rightarrow 5^1(-1)^+ $            &$ 516.80 \pm 0.22 $   &  $ 0.0046 \pm 0.0008 $ & max      \\	 
$ 2^10^+ \rightarrow 2^1(-1)^+ $            &$ 5285   \pm 3    $   &  $  0.021 \pm 0.003  $ & max      \\	 
$ 2^10^+ \rightarrow 2^1(-1)^+ $            &$ 4812   \pm 30   $   &  $  0.11  \pm 0.01   $ & min      \\	 
$ 2^10^+ \rightarrow 3^1(-1)^+ $            &$ 4125   \pm 3    $   &  $  0.021 \pm 0.003  $ & max      \\	 
$ 2^10^+ \rightarrow 3^1(-1)^+ $            &$ 3417   \pm 10   $   &  $  0.183 \pm 0.005  $ & min      \\	 
$ 2^10^+ \rightarrow 3^1(-1)^+ $            &$ 3795.8 \pm 0.5  $   &  $  0.802 \pm 0.008  $ & max      \\	 
$ 2^10^+ \rightarrow 4^1(-1)^+ $            &$ 4006.2 \pm 0.8  $   &  $ 0.004 \pm 0.001   $ & max      \\	 
$ 2^10^+ \rightarrow 4^1(-1)^+ $            &$ 3015   \pm 10   $   &  $  0.20 \pm 0.01    $ & min      \\	 
$ 2^10^+ \rightarrow 4^1(-1)^+ $            &$ 3177.7 \pm 1.0  $   &  $ 0.661 \pm 0.005   $ & max      \\	 
$ 2^10^+ \rightarrow 5^1(-1)^+ $            &$ 3651.7 \pm 1.4  $   &  $ 0.0050 \pm 0.0005 $ & max      \\	 
$ 2^10^+ \rightarrow 5^1(-1)^+ $            &$ 2825   \pm 3    $   &  $  0.22 \pm 0.01    $ & min      \\	 
$ 2^10^+ \rightarrow 5^1(-1)^+ $            &$ 2950.4 \pm 1.7  $   &  $ 0.637 \pm 0.006   $ & max      \\	 
$ 3^10^+ \rightarrow 3^1(-1)^+ $            &$ 22030  \pm  65  $   &  $ 0.034 \pm 0.001   $ & max      \\
$ 3^10^+ \rightarrow 3^1(-1)^+ $            &$ 12634.5\pm 5.4  $   &  $ 0.177 \pm 0.002   $ & min      \\
$ 3^10^+ \rightarrow 4^1(-1)^+ $            &$ 15809  \pm 40   $   &  $0.00576 \pm 0.00008$ & max      \\
$ 3^10^+ \rightarrow 4^1(-1)^+ $            &$ 8270   \pm 150  $   &  $  0.195 \pm 0.005  $ & min      \\
$ 3^10^+ \rightarrow 4^1(-1)^+ $            &$ 11765  \pm 6    $   &  $  1.43 \pm 0.02    $ & max      \\
$ 3^10^+ \rightarrow 5^1(-1)^+ $            &$ 11533  \pm  26  $   &  $  0.011 \pm 0.002  $ & max      \\	
$ 3^10^+ \rightarrow 5^1(-1)^+ $            &$ 7170   \pm 20   $   &  $  0.210 \pm 0.005  $ & min      \\	
$ 3^10^+ \rightarrow 5^1(-1)^+ $            &$ 9004.4 \pm 5.5  $   &  $  1.08 \pm 0.02    $ & max      \\	
$ 4^10^+ \rightarrow 4^1(-1)^+ $            &$ 19840  \pm  290 $   &  $  0.004 \pm 0.002  $ & max      \\
$ 4^10^+ \rightarrow 4^1(-1)^+ $            &$ 16190  \pm  170 $   &  $  0.024 \pm 0.003  $ & min      \\
$ 4^10^+ \rightarrow 4^1(-1)^+ $            &$ 48900  \pm  1400$   &  $  0.074 \pm 0.009  $ & max      \\
$ 4^10^+ \rightarrow 4^1(-1)^+ $            &$ 26100  \pm  1700$   &  $  0.194 \pm 0.004  $ & min      \\
$ 4^10^+ \rightarrow 5^1(-1)^+ $            &$ 13283  \pm 8    $   &  $ 0.0065 \pm 0.0004 $ & max      \\
$ 4^10^+ \rightarrow 5^1(-1)^+ $            &$ 10490  \pm 550  $   &  $ 0.0285 \pm 0.0005 $ & min      \\
$ 4^10^+ \rightarrow 5^1(-1)^+ $            &$ 21520  \pm 220  $   &  $ 0.092 \pm 0.003   $ & max      \\
$ 4^10^+ \rightarrow 5^1(-1)^+ $            &$ 17100  \pm 230  $   &  $ 0.210 \pm 0.003   $ & max      \\
\end{tabular}
\end{minipage}
\end{table}

\begin{table}
\begin{minipage}{11.0cm}
\caption{Same as Table \ref{statsing} but for the triplet $|\Delta M| = 1$  transitions 
within the subspace of positive $z$-parity, $\mu^3(-1)^+ \rightarrow \nu^30^+$ 
for $\mu = 1,...,5,\;\nu = 1,...,5$. }
\vspace{1mm}

\begin{tabular}{l   l   l   l}\hline\hline
           Component                        & wavelength/A        & position $B/a.u.$   &   max/min      \\ 
        $\nu^{2S+1}M^{(-1)^{\Pi_z}}$        &                     &                     &                \\ \hline
$ 1^30^+ \rightarrow 2^3(-1)^+ $            & $ 4058.6\pm 0.4$    & $ 0.0214 \pm 0.0001    $ & max       \\
$ 1^30^+ \rightarrow 2^3(-1)^+ $            & $ 3580  \pm 2  $    & $    0.178 \pm 0.002   $ & min       \\
$ 1^30^+ \rightarrow 3^3(-1)^+ $            & $ 3258.1\pm 0.25$   & $ 0.020 \pm 0.001      $ & max       \\
$ 1^30^+ \rightarrow 3^3(-1)^+ $            & $ 2615  \pm 5  $    & $ 0.25  \pm 0.02       $ & min       \\
$ 1^30^+ \rightarrow 3^3(-1)^+ $            & $ 3090.1\pm 0.3$    & $ 1.757 \pm 0.004      $ & max       \\
$ 1^30^+ \rightarrow 4^3(-1)^+ $            & $ 3210  \pm  2 $    & $ 0.00576 \pm 0.00008  $ & max       \\
$ 1^30^+ \rightarrow 4^3(-1)^+ $            & $ 2336  \pm 5  $    & $ 0.28  \pm 0.01       $ & min       \\
$ 1^30^+ \rightarrow 4^3(-1)^+ $            & $ 2600.4\pm 0.5$    & $ 1.460 \pm 0.009      $ & max       \\
$ 1^30^+ \rightarrow 5^3(-1)^+ $            & $ 2959.2\pm 0.7$    & $ 0.0047 \pm 0.001     $ & max       \\
$ 1^30^+ \rightarrow 5^3(-1)^+ $            & $ 2208  \pm 3  $    & $  0.299 \pm 0.005     $ & min       \\
$ 1^30^+ \rightarrow 5^3(-1)^+ $            & $ 2428.1\pm 0.5$    & $ 1.379  \pm 0.009     $ & max       \\
$ 2^30^+ \rightarrow 3^3(-1)^+ $            & $ 17690 \pm  80$    & $ 0.047 \pm 0.002      $ & max       \\
$ 2^30^+ \rightarrow 3^3(-1)^+ $            & $ 11058.2\pm 2.0$   & $ 0.238 \pm 0.005      $ & min       \\
$ 2^30^+ \rightarrow 4^3(-1)^+ $            & $ 12990 \pm 32 $    & $ 0.0065 \pm 0.0004    $ & max       \\
$ 2^30^+ \rightarrow 4^3(-1)^+ $            & $ 7410  \pm 40 $    & $   0.27 \pm 0.02      $ & min       \\
$ 2^30^+ \rightarrow 4^3(-1)^+ $            & $ 10485 \pm  9 $    & $   2.93 \pm 0.07      $ & max       \\
$ 2^30^+ \rightarrow 5^3(-1)^+ $            & $ 9711  \pm 13 $    & $  0.013 \pm 0.003     $ & max       \\
$ 2^30^+ \rightarrow 5^3(-1)^+ $            & $ 6250  \pm 40 $    & $  0.29  \pm 0.01      $ & min       \\
$ 2^30^+ \rightarrow 5^3(-1)^+ $            & $ 8031.8\pm 1.7$    & $   2.14 \pm 0.08      $ & max       \\
$ 3^30^+ \rightarrow 4^3(-1)^+ $            & $ 20450 \pm 120 $   & $ 0.007 \pm 0.001      $ & max       \\
$ 3^30^+ \rightarrow 4^3(-1)^+ $            & $ 14850 \pm 300 $   & $ 0.035 \pm 0.003      $ & min       \\
$ 3^30^+ \rightarrow 4^3(-1)^+ $            & $ 37220 \pm 680 $   & $ 0.094 \pm 0.003      $ & max       \\
$ 3^30^+ \rightarrow 4^3(-1)^+ $            & $ 26080 \pm 250 $   & $ 0.281 \pm 0.009      $ & min       \\
$ 3^30^+ \rightarrow 5^3(-1)^+ $            & $ 13320 \pm 3   $   & $ 0.0078 \pm 0.0009    $ & max       \\
$ 3^30^+ \rightarrow 5^3(-1)^+ $            & $ 11150 \pm 560 $   & $ 0.032 \pm 0.003      $ & min       \\
$ 3^30^+ \rightarrow 5^3(-1)^+ $            & $ 20300 \pm 1000$   & $ 0.114 \pm 0.007      $ & max       \\
$ 3^30^+ \rightarrow 5^3(-1)^+ $            & $ 16030 \pm 210 $   & $ 0.29 \pm 0.01        $ & min       \\
$ 3^30^+ \rightarrow 5^3(-1)^+ $            & $ 24864 \pm 11  $   & $ 4.19 \pm 0.08        $ & max       \\
$ 4^30^+ \rightarrow 4^3(-1)^+ $            & $ 296800\pm 8500$   & $ 0.009  \pm 0.001     $ & max       \\
$ 4^30^+ \rightarrow 4^3(-1)^+ $            & $ 297000\pm 8500$   & $ 0.26 \pm 0.02        $ & max       \\
$ 4^30^+ \rightarrow 5^3(-1)^+ $            & $ 42500 \pm 1300$   & $ 0.017 \pm 0.002      $ & max       \\
$ 4^30^+ \rightarrow 5^3(-1)^+ $            & $ 82600 \pm 3300$   & $ 0.15 \pm 0.02        $ & max       \\
$ 4^30^+ \rightarrow 5^3(-1)^+ $            & $ 46400 \pm 2000$   & $ 0.29 \pm 0.02        $ & min       \\
$ 4^30^+ \rightarrow 5^3(-1)^+ $            & $ 240400\pm 1400$   & $ 8.3 \pm 0.5          $ & max       \\
\end{tabular}
\end{minipage}
\end{table}

\begin{table}
\begin{minipage}{11.0cm}
\caption{Overview over all of the stationary points found in the singlet $|\Delta M| = 1$ transitions 
within the subspace of negative $z$-parity, $\mu^1(-1)^- \rightarrow \nu^10^-$ for $\mu = 1,...,5,\;\nu = 1,...,5$.}
\vspace{1mm}
\label{statsing2}
\begin{tabular}{l   l   l   l}\hline\hline
           Component                        & wavelength/A         & position $B/a.u.$    & max/min  \\ 
$\nu^{2S+1}M^{(-1)^{\Pi_z}} $               &                      &                      &          \\ \hline
$2^1(-1)^-  \rightarrow  1^1 0^-$           & $5056  \pm 22$       & $0.0088 \pm 0.0002$  &  max     \\	 
$3^1(-1)^-  \rightarrow  1^1 0^-$           & $4477  \pm 18$       & $0.0070 \pm 0.0003$  &  max     \\	 
$3^1(-1)^-  \rightarrow  2^1 0^-$           & $13875 \pm 42$       & $0.0105 \pm  0.0001$ &  max     \\	 
$3^1(-1)^-  \rightarrow  3^1 0^-$           & $56300 \pm 1500$     & $0.0139 \pm  0.0015$ &  max     \\	 
$3^1(-1)^-  \rightarrow  3^1 0^-$           & $50841 \pm 100$      & $0.0387 \pm  0.0035$ &  min     \\	 
$4^1(-1)^-  \rightarrow  1^1 0^-$           & $4442  \pm 20$       & $0.0022 \pm 0.0006$  &  max     \\	 
$4^1(-1)^-  \rightarrow  2^1 0^-$           & $13327 \pm 75$       & $0.0025 \pm 0.0007$  &  max     \\	 
$4^1(-1)^-  \rightarrow  3^1 0^-$           & $44050 \pm 570$      & $0.0027 \pm 0.0009$  &  max     \\	 
$5^1(-1)^-  \rightarrow  1^1 0^-$           & $4188  \pm 16$       & $0.0030 \pm 0.0008$  &  max     \\
$5^1(-1)^-  \rightarrow  2^1 0^-$           & $11292 \pm 32$       & $0.0035 \pm 0.001$   &  max     \\
$5^1(-1)^-  \rightarrow  3^1 0^-$           & $27640 \pm 30$       & $0.004 \pm 0.001$    &  max     \\
$5^1(-1)^-  \rightarrow  4^1 0^-$           & $48860 \pm 280$      & $0.195 \pm 0.003$    &  max     \\
$5^1(-1)^-  \rightarrow  4^1 0^-$           & $45700 \pm 300$      & $5.2 \pm 0.2$        &  min     \\
$5^1(-1)^-  \rightarrow  5^1 0^-$           & $98980 \pm 600$      & $0.0065 \pm 0.0003$  &  max     \\	 
$5^1(-1)^-  \rightarrow  5^1 0^-$           & $83470 \pm 420$      & $0.0221 \pm 0.0009$  &  min     \\	 
\end{tabular}
\end{minipage}
\end{table}

\begin{table}
\begin{minipage}{11.0cm}
\caption{Same as Table \ref{statsing2} but for the triplet $|\Delta M| = 1$ 
within the subspace of negative $z$-parity, 
transitions $\mu^3(-1)^- \rightarrow \nu^30^-$ 
for $\mu = 1,...,5,\;\nu = 1,...,5$. }
\vspace{1mm}

\begin{tabular}{l   l   l   l}\hline\hline
           Component                        & wavelength/A        & position $B/a.u.$   &   max/min     \\ 
        $\nu^{2S+1}M^{(-1)^{\Pi_z}}$        &                     &                     &               \\ \hline
$1^3(-1)^-  \rightarrow  1^3 0^-$           & $7143  \pm 30$      & $0.1103 \pm 0.0007$ & max           \\	 
$2^3(-1)^-  \rightarrow  1^3 0^-$           & $4574  \pm 15$      & $0.0087 \pm 0.0003$ & max           \\	 
$3^3(-1)^-  \rightarrow  1^3 0^-$           & $4095  \pm 10$      & $0.0070 \pm 0.0003$ & max           \\	 
$3^3(-1)^-  \rightarrow  2^3 0^-$           & $12680 \pm 22$      & $0.00965\pm 0.00006$& max           \\	 
$3^3(-1)^-  \rightarrow  3^3 0^-$           & $51500 \pm 1300$    & $0.013 \pm 0.001$   & max           \\	 
$3^3(-1)^-  \rightarrow  3^3 0^-$           & $43530 \pm 400$     & $0.046 \pm 0.004$   & min           \\	 
$4^3(-1)^-  \rightarrow  1^3 0^-$           & $4063  \pm 12 $     & $0.0024 \pm 0.0003$ & max           \\	 
$4^3(-1)^-  \rightarrow  2^3 0^-$           & $12263 \pm 25$      & $0.00224\pm 0.00007$& max           \\
$4^3(-1)^-  \rightarrow  3^3 0^-$           & $40265 \pm 110$     & $0.00264\pm 0.00001$& max           \\
$4^3(-1)^-  \rightarrow  3^3 0^-$           & $23021 \pm 40$      & $0.079 \pm 0.007$   & min           \\
$4^3(-1)^-  \rightarrow  3^3 0^-$           & $23532 \pm 50$      & $0.17 \pm 0.01$     & max           \\
$5^3(-1)^-  \rightarrow  1^3 0^-$           & $3850  \pm 9$       & $0.0033 \pm 0.0009$ & max           \\
$5^3(-1)^-  \rightarrow  2^3 0^-$           & $10505 \pm 20$      & $0.0035 \pm 0.0007$ & max           \\	 
$5^3(-1)^-  \rightarrow  3^3 0^-$           & $26225 \pm 70$      & $0.0050 \pm 0.0005$ & max           \\	 
$5^3(-1)^-  \rightarrow  4^3 0^-$           & $44662 \pm 30$      & $0.24 \pm 0.02$     & max           \\
$5^3(-1)^-  \rightarrow  5^3 0^-$           & $91500 \pm 1200$    & $0.0062 \pm 0.0003$ & max           \\
$5^3(-1)^-  \rightarrow  5^3 0^-$           & $2950  \pm 530 $    & $0.022 \pm 0.001$   & min           \\
\end{tabular}
\end{minipage}
\end{table}

\newpage